\documentclass[a4paper,12pt]{article}
\usepackage{graphicx}
\usepackage{amsmath}
\usepackage{amssymb}
\usepackage{multirow}
\usepackage{float} 

\begin{document}

\begin{flushright}
preprint SHEP-10-37\\
\today
\end{flushright}
\vspace*{1.0truecm}

\begin{center}
{\large\bf Low mass Higgs signals at the LHC\\[0.15cm]
in the Next-to-Minimal Supersymmetric Standard Model}\\
\vspace*{1.0truecm}
{\large M. M. Almarashi and S. Moretti}\\
\vspace*{0.5truecm}
{\it School of Physics and Astronomy, \\
 University of Southampton, Southampton, SO17 1BJ, UK}
\end{center}

\vspace*{2.0truecm}
\begin{center}
\begin{abstract}
\noindent 

In the NMSSM, because of introducing a complex singlet superfield, the lightest CP-odd Higgs boson, $a_1$, can be
a singlet-like state with a tiny doublet component in large regions of parameter space. In this paper, we examine
the discovery potential of $a_1$ produced in association with a bottom-antibottom pair at the LHC through
$\tau^+\tau^-$ and $\gamma\gamma$ decay modes. It is shown that an $a_1$ with mass $\leq M_Z$ can be 
extracted from the SM backgrounds by using the $\tau^+\tau^-$ decay channel, a possibility precluded to
the MSSM. In contrast, the $\gamma\gamma$ decay mode is overwhelmed by backgrounds despite the fact that
the branching ratio of this mode can reach unity when $a_1$ is a pure singlet.

\end{abstract}
\end{center}

\newcommand{\nn}{\nonumber}
\section{Introduction}
\label{sect:intro}

In the Next-to-Minimal Supersymmetric Standard Model (NMSSM) \cite{review}, 
the soft Supersymmetry (SUSY)-breaking Higgs sector is described by the
Lagrangian contribution
\begin{equation}
V_{\rm NMSSM}=m_{H_u}^2|H_u|^2+m_{H_d}^2|H_d|^2+m_{S}^2|S|^2
             +\left(\lambda A_\lambda S H_u H_d + \frac{1}{3}\kappa A_\kappa S^3 + {\rm h.c.}\right),
\end{equation}
where $H_u$ and $H_d$ are the Higgs doublet fields, $S$ the singlet one, $\lambda$ and $\kappa$ Yukawa
couplings while $A_\lambda$ and $A_\kappa$ are dimensionful parameters of order $M_{\rm{SUSY}}$,
the typical SUSY mass scale. 

As a result of the introduction of an extra complex singlet scalar
field, which only couples to the two MSSM-type Higgs doublets, the
Higgs sector of the NMSSM comprises of a total of seven mass
eigenstates: a charged pair $h^\pm$, three CP-even Higgses
$h_{1,2,3}$ ($m_{h_1}<m_{h_2}<m_{h_3}$) and two CP-odd
Higgses $a_{1,2}$ ($m_{a_1}<m_{a_2}$). Consequently, Higgs
phenomenology in the NMSSM may be plausibly different from that of the
MSSM and extremely rich of new signals.\\

For a start, the sum of the squares of the two lightest scalar
Higgs boson masses is given by \cite{MNZ}:
\begin{eqnarray}
m_{h_1}^2 + m_{h_2}^2 \approx M_Z^2 + \frac{1}{2}\kappa <S>(4\kappa <S>+\sqrt{2}A_\kappa)
\end{eqnarray}
with $m_{h_1}^2\leq m_{h_2}^2$. The last expression  
can be translated into a modified upper bound of the $h_1$ mass as \cite{MNZ}
\begin{eqnarray}
m_{h_1}^2 \lesssim {\rm min}\{M_Z^2,\frac{1}{2}\kappa <S>(4\kappa
<S>+\sqrt{2}A_\kappa)\}, 
\end{eqnarray}
so that the upper bound on the NMSSM lightest Higgs boson mass is higher than the corresponding bound in the MSSM.
Furthermore, as the higher order corrections are similar to those in
the MSSM, also in higher orders the upper bound on the lightest Higgs boson mass
remains different in the NMSSM with respect to the MSSM, reaching 140 GeV or so, for maximal
stop mixing and $\tan\beta=2$ \cite{upper,Cyril} (a configuration indeed excluded
in the MSSM by LEP data). More in general, the
`little fine tuning problem', resulting in LEP failing
to detect a light CP-even Higgs boson, predicted over most
of the MSSM parameter space, is much attenuated in the NMSSM,
either because a SM-like Higgs can decay dominantly into $a_1a_1$ 
\cite{LEPescape} or because the mixing among more numerous CP-even or CP-odd
Higgs fields enables light mass states being produced at LEP
or Tevatron yet, they can remain undetected because of their reduced couplings
to $Z$ bosons \cite{Cyril}. (See Refs.~\cite{NMSSM-Points} and \cite{NMSSM-Benchmarks} 
for benchmark points with very light CP-odd Higgs bosons in both the unconstrained and 
constrained NMSSM parameter space, respectively).  
  
As for present accelerator machines, chiefly the Tevatron at FNAL 
and the Large Hadron Collider (LHC) at CERN,  
quite some work has been dedicated to probing the NMSSM
Higgs sector over recent years. Primarily, there have been attempts
to extend the so-called `No-lose theorem' of the MSSM -- stating 
 that at least one MSSM Higgs boson should be observed via
the usual SM-like production and decay channels at the LHC 
 throughout the entire MSSM parameter space \cite{NoLoseMSSM} --
to the case of the NMSSM \cite{NMSSM-Points,NoLoseNMSSM1,Shobig1,Shobig2,Erice}. From this perspective,
it was realised that at least one NMSSM Higgs boson should remain observable 
at the LHC over the NMSSM parameter space that does not allow any Higgs-to-Higgs 
decay. However, when the only light non-singlet (and, therefore, potentially visible) CP-even
Higgs boson, $h_1$ or $h_2$, decays mainly to two very light 
CP-odd Higgs bosons, $h\to a_1 a_1$, one
may not have a Higgs signal of statistical significance at the LHC \cite{dirk}. In fact, 
further violations to the theorem may well occur if
one enables Higgs-to-SUSY decays (e.g., into neutralino pairs, yielding invisible Higgs signals).

While there is no definitive evidence on whether a `No-lose theorem' can be proved for the NMSSM,
we are here also concerned with an orthogonal approach. We asked ourselves if a, so to say,
`More-to-gain theorem' can be formulated in the NMSSM. That is, whether there exist regions
of the NMSSM parameter space where  more and/or different Higgs
states of the NMSSM are visible at the LHC than those available within the MSSM. 
In our attempt to overview such a possibility, we consider here the case
of the di-photon and $\tau^+\tau^-$  decay channels of a light neutral Higgs boson.  The first mode can successfully
be probed in the MSSM, but limitedly to the case of the lightest CP-even Higgs boson and for masses
of order 100--130 GeV or so. We will show that in the NMSSM there exist regions of its parameter
space where one can potentially have a sizable di-photon signal
from a Higgs state of different nature (CP-odd instead of CP-even) and with much smaller mass (down to 10
GeV or so), owing in large part to an increased BR$(a_1\to\gamma\gamma)$. Such a possibility emerges in
the NMSSM due to the fact that such a CP-odd Higgs state has a predominant 
singlet component and a very weak doublet one. As a
consequence, all partial decay widths are heavily suppressed as they
employ only the doublet component, except one: the $\gamma\gamma$ partial decay
width. This comes from the fact that the $a_1\tilde\chi^+\tilde\chi^-$
coupling is not suppressed, as it is generated through the $\lambda H_1 H_2 S$ Lagrangian
term and therefore implies no small mixing. Although the direct decay 
$a_1$ $\to$ $\tilde\chi^+\tilde\chi^-$ is forbidden, the aforementioned
coupling participates in the $a_1\gamma\gamma$ effective coupling. 
The $\tau^+\tau^-$ channel is used in the MSSM as a search channel of rather heavy and degenerate
CP-even and CP-odd states and its exploitation has not been proved at very low masses, say, below $M_Z$.
We will show here that the $\tau$-pair decay can be very relevant for such a mass interval, in the case
of an $a_1$ state of the NMSSM. The extraction of either channel for low Higgs masses would then unmistakably
point to the existence of non-minimal SUSY Higgs sector.  
We build on the results presented in \cite{Shobig1} (see also \cite{CPNSH}), where several such signals (for rather heavy Higgs 
states though) were established in some (complementary) regions of the NMSSM parameter space.

The plan of this paper is as follows. In Sec.~\ref{sect:scan} we describe the parameter space scans
performed in order to isolate the combinations of NMSSM inputs that can give rise to the described
phenomenology. In Sec.~\ref{sect:rates} we discuss typical event rates, highlighting the relevance
of Higgs production in association with bottom quark pairs as most favourable production mode.
In Sec.~\ref{sect:S2B} we describe the signal-to-background analysis performed and introduce
some benchmark points where an $a_1$ signal can be extracted
(albeit limitedly to the $\tau^+\tau^-$ decay mode). Finally, in Sec.~\ref{sect:summa} we summarise and
conclude.  

\section{\large Parameter Space Scan}
\label{sect:scan}

For a general study of the NMSSM Higgs sector,
we used here the up-to-date publicly available fortran package NMSSMTools developed in Refs.~\cite{NMHDECAY,NMSSMTools}\footnote
{We have used NMSSMTools$\_$2.3.1.}. This package
computes the masses, couplings and decay widths of all the Higgs
bosons of the NMSSM in terms of its parameters at the Electro-Weak (EW)
scale. The NMSSMTools also takes into account theoretical as well as
experimental constraints from negative Higgs searches at LEP \cite{LEP} and the Tevatron\footnote{Speculations
of an excess at LEP which could be attributed to NMSSM Higgs bosons are found in \cite{excess}.},
along with the unconventional channels relevant for the NMSSM.

For our purpose, instead of postulating unification, and without taking into account the 
SUSY breaking mechanism, we fixed the soft SUSY breaking terms to very high values, 
so that they have little or no contribution to the outputs of the parameter scans. 
Consequently, we are left with the aforementioned six free parameters.
Our  parameter space is in particular defined through the Yukawa couplings $\lambda$ and
$\kappa$, the soft trilinear terms $A_\lambda$ and $A_\kappa$, plus 
tan$\beta$ (the ratio of the 
Vacuum Expectation Values (VEVs) of the two Higgs doublets) and $\mu_{\rm eff} = \lambda\langle S\rangle$
(where $\langle S\rangle$ is the VEV of the Higgs singlet). In our numerical analyses we have taken 
$m^{\rm pole}_b=5$ GeV and $m^{\rm pole}_t=171.4$ GeV 
for the bottom- and top-quark mass, respectively.
 
 We have used the NMHDECAY code to scan over the NMSSM
parameter space defined through the six parameters already discussed 
taken in the following intervals:
\begin{center}
$\lambda$ : 0.0001 -- 0.7,\phantom{aa} $\kappa$ : 0 --
0.65,\phantom{aa} $\tan\beta$ : 1.6 -- 54,\\ $\mu$ : 100 -- 1000 GeV,\phantom{aa} $A_{\lambda}$ : $-$1000 -- +1000 GeV,\phantom{aa} $A_{\kappa}$ :$-$10 -- 0.\\
\end{center}
\noindent
(Notice that our aim is exploring the parameter space which has very low $m_{a_1}$ and one way to do that is by choosing $A_\kappa$ small,
in which case its negative value is preferred \cite{MNZ}).

Remaining soft terms which are fixed in the scan include:\\
$\bullet\phantom{a}m_{Q_3} = m_{U_3} = m_{D_3} = m_{L_3} = m_{E_3} = 1$ TeV, \\
$\bullet\phantom{a}A_{U_3} = A_{D_3} = A_{E_3} = 1.2$ TeV,\\
$\bullet\phantom{a}m_Q = m_U = m_D = m_L = m_E = 1$ TeV,\\
$\bullet\phantom{a} M_1 = M_2 = M_3 = 1.5$ TeV.\\

\noindent
As intimated, we have fixed soft term parameters at the TeV scale to minimise their contributions to parameter space
outputs but changing values of some of those parameters such as $A_{U_3}$ could decrease or increase the number of successful
points emerging from the NMSSMTools scans but without a significant impact on the $m_{a_1}$ distribution.
Also, notice that the sfermion mass parameters and the $SU(2)$ gaugino mass parameter, $M_2$, play 
crucial roles in constraining $\tan\beta$. Lowering values of those parameters allow less values of $\tan\beta$ to 
pass experimental and theoretical constraints, however, this is a less interesting region of the NMSSM parameter
space for our analysis, as our Higgs production mode is only relevant at large values of this parameter. The effect
of heavy gaugino mass parameters on the output of parameter space, in particular $m_{a1}$, would be small except
for $M_2$ through its effect on $tan\beta$.

In line with the assumptions made in \cite{NMSSM-Points,NoLoseNMSSM1}, the allowed decay 
modes for neutral NMSSM Higgs bosons are\footnote{Here, we use the label
$h(a)$ to signify any of the neutral CP-even(odd) Higgs bosons of the NMSSM.}:
\begin{eqnarray}
h,a\rightarrow gg,\phantom{aaa} h,a\rightarrow \mu^+\mu^-,
&&h,a\rightarrow\tau^+\tau^-,\phantom{aaa}h,a\rightarrow
b\bar b,\phantom{aaa}h,a\rightarrow t\bar t,\nn \\ h,a\rightarrow
s\bar s,\phantom{aaa}h,a\rightarrow
c\bar c,&&h\rightarrow W^+W^-,\phantom{aaa}h\rightarrow ZZ,\nn \\
h,a\rightarrow\gamma\gamma,\phantom{aaa}h,a\rightarrow
Z\gamma,&&h,a\rightarrow {\rm Higgses},\phantom{aaa}h,a\rightarrow
{\rm sparticles}. \nn
\end{eqnarray}
(Notice that for the CP-odd Higgses, the decay into vector
boson pairs is not allowed due to CP-conservation). Here, 
`Higgses' refers to any final state
involving all possible combination of two Higgs bosons (neutral and/or charged)
or of one Higgs boson and a gauge vector.

We have performed our scan over 10 million of randomly
selected points in the specified parameter space. The output, as
stated earlier, contains masses, Branching Ratios (BRs) and couplings of
the NMSSM Higgses, for all the points which are not forbidden by
the various experimental and theoretical constraints. The points which violate the constraints
are eliminated. 

\section{\large Inclusive Event Rates}
\label{sect:rates}

The surviving data points are then used to determine the
cross sections for NMSSM Higgs hadro-production by using CalcHEP \cite{CalcHEP}\footnote{We adopt herein
CTEQ6L \cite{cteq} as parton distribution functions, with scale $Q=\sqrt{\hat{s}}$, the centre-of-mass energy
at parton level, for all processes computed.}, wherein some new modules 
have been implemented for this purpose \cite{Almarashi}. As the SUSY mass scale 
has been arbitrarily set well above the EW one (see above), 
the NMSSM Higgs production modes
exploitable in simulations at the LHC are those involving couplings to
heavy ordinary matter only, i.e., (hereafter, $V=W^\pm,Z$ and $Q=b,t$)
for neutral Higgs production (where the last two channels are only allowed
for CP-even Higgs production):
\vspace*{-0.2cm}
$$gg\to {\rm{Higgs}}~({\rm{gluon-fusion,~via~heavy-quark~loops}}),$$
$$q\bar q,gg\to Q\bar Q~{\rm{Higgs}}~({\rm{heavy-quark~associated~production}}),$$
$$qq\to qq V^{*}V^{*}\to qq~{\rm{Higgs}}~({\rm{Vector~Boson~Fusion~(VBF)}}),$$
$$q\bar q\to V~{\rm{Higgs}}~({\rm{Higgs-strahlung}}).$$
(These are the so-called `direct' Higgs production modes). Here, 
`Higgs' refers to any possible neutral Higgs boson. We can however anticipate that, for the
purpose of extracting an $a_1\to\gamma\gamma$  or $a_1\to \tau^+\tau^-$ resonance, only the second channel turns 
out to be useful and limitedly to the case $Q=b$. This is because the associated $b\bar b$ pairs can be (vertex)
tagged, thus offering a useful handle for background rejection. The case $Q=t$ in fact gives too small 
cross section.
The gluon fusion channel is instead burdened by huge Standard Model
(SM) backgrounds ($q\bar q,gg\to \gamma\gamma$ as well as jets 
mis-identified as photons in the $\gamma\gamma$ mode and Drell-Yan and di-jet production in the 
$\tau^+\tau^-$ channel). Higgs-strahlung and VBF also suffer from large SM backgrounds, though their main
drawback are rather smaller production rates in comparison (as they are EW processes).\\

In the NMSSM, $a_1$ state is a composition of the usual doublet component of the CP-odd MSSM Higgs boson, $a_{\rm MSSM}$, and
the new singlet component, $a_{\rm S}$, coming from the singlet superfield of the NMSSM. This can be written as \cite{excess}:
\begin{equation}
 a_1=a_{\rm MSSM}\cos\theta_A+a_{\rm S}\sin\theta_A.
\end{equation}
For very small values of $A_k$, the lightest CP-odd Higgs, $a_1$, is mostly singlet-like
with a tiny doublet component, i.e. the mixing angle $\cos\theta_A$ is small, see
top-pane of Fig. 1, which shows the relation between $m_{a_1}$ and $\cos\theta_A$. If $a_1$ is highly singlet, $\cos\theta \sim 0$, 
then BR$(a_1\to \gamma\gamma)$ can reach unity as shown in the bottom-pane of this figure.\\ 

To good approximation, $m_{a_1}$ can be written in the NMSSM as \cite{excess}:
\begin{equation} 
 m^2_{a_1}=-3\frac{\kappa A_\kappa \mu_{\rm eff}}{\lambda}\sin^2\theta_A+\frac{9A_\lambda \mu_{\rm eff}}{2\sin2\beta}\cos^2\theta_A.
\end{equation}
The first term of the this expression is dominant especially at large $\tan\beta$. Also, it is clear that
a combination of all tree level Higgs sector parameters jointly affects $m_{a_1}$ in general.

As an initial step towards the analysis of the data, we have computed $m_{a_1}$ against each of
the six tree level Higgs sector parameters of the NMSSM. Fig. 2 presents the results of our scan, these series of plots
illustrating the distribution of $m_{a_1}$ over the six parameters and as a function of BR$(a_1\to\gamma\gamma)$
and BR$(a_1\to \tau^+\tau^-)$. In the parameter space adopted for this analysis we have noticed that the large $\tan\beta$ and 
small $\mu_{\rm eff}$ (and, to some extent, also $\lambda$) region
is the one most compatible with current theoretical and experimental constraints, though this conclusion should not be
generalised to the entire parameter space. Herein, it is also obvious that $m_{a_1}$ increases
by increasing $\kappa$ and $-A_\kappa$ whereas it decreases by increasing $A_\lambda$. Moreover, from a closer look at Fig. 2,
it is clear that the BR$(a_1\to\gamma\gamma)$ can be very large, indeed dominant, and $m_{a_1}$ values 
in the region 50 to 100 GeV maximise that rate. Also, notice that BR$(a_1\to \tau^+\tau^-)$
reaches about 10\% in most of the parameter space which has $m_{a_1}\geq 10$ GeV, in which case the $a_1$ decay into 
$b\bar b$ is open and dominant. Yet also notice that there is a small region with very low $m_{a_1}$ values, $m_{a_1}< 10 $ GeV, 
yielding a large BR$(a_1\to \tau^+\tau^-)$, greater than $90\%$, 
as shown in the bottom-right corner of the bottom-right pane of this figure.
The latter region occurs when $a_1\to  b\bar b$ is closed, in which case BR$(a_1\to \tau^+\tau^-)$ is dominant
compared to $c\bar c$, $\mu^+\mu^-$, etc\footnote{Notice that the mass region below the $b\bar b$ threshold is severely 
constrained, see, e.g., Ref. \cite{Lebedev}
(and references therein).}.

Fig. 3 shows the distribution of event rates, $\sigma(gg\to b\bar b {a_1})~{\rm BR}(a_1\to \gamma\gamma)$ and
$\sigma(gg\to b\bar b {a_1})~{\rm BR}(a_1\to \tau^+\tau^-)$ as functions of $m_{a_1}$, BR of the corresponding channel and
$\tan\beta$. As expected, the inclusive cross section decreases with increasing $m_{a_1}$, 
see the top two panes of this figure. Although the BR$(a_1\to \gamma\gamma)$ can be dominant over a sizable expanse of the NMSSM parameter space 
(middle-left pane), 
the latter does not correspond to the region that maximises the yield of $\sigma(gg\to b\bar b {a_1})~{\rm BR}(a_1\to \gamma\gamma)$, 
as the maximum
of the latter occurs for BRs in the region of a few $10^{-5}$ to $10^{-4}$. Therefore, one cannot 
take full advantage of the phenomenon described in the introduction, with respect to the singlet nature of the
$a_1$ state entering the $a_1\tilde\chi^+\tilde\chi^-$ coupling, as the doublet component (necessary to enable a 
large $a_1b\bar b$ coupling at production level) plays a stronger role in comparison. The tension between the two 
trends is such that the cross section times BR rates are less than 100 fb. The outlook for the $\tau^+\tau^-$ case 
is much brighter, as corresponding signal rates are at nb level for 
BR$(a_1\to \tau^+\tau^-)\approx 0.1$ or even 10 nb for  BR$(a_1\to \tau^+\tau^-)\approx 1$, see middle-right
pane of Fig. 3. Also notice that such large rates naturally occur for any $m_{a_1}$ 
in the allowed interval (see top-right pane of this figure).

Incidentally, notice in the case of both decay channels that not only the density of NMSSM 
parameter configurations is larger as $\tan\beta$ grows\footnote{Again,
notice that for a more general choice of the range of $A_\kappa$ and values of the soft SUSY breaking parameters this would not necessarily hold.} 
but also the event rates are maximal
at large values of this parameter (see bottom-left and bottom-right panes of Fig. 3), 
thereby confirming what we intimated at the beginning of this
section about the relevance of the $q\bar q,gg\to b\bar b a_1$ production mode (whose cross
section is essentially proportional to $\tan^2\beta$).

In the NMSSM, there is a large area of parameter space where one Higgs state can decay into two, e.g., $h_1\to a_1a_1$: see
Fig. 4.  As it is clear from the left-pane of this figure, the majority of points generated here have $m_{h1}>110$ GeV and 
$m_{a_1}<55$ GeV, thereby allowing the possibility of $h_1\to a_1a_1$ decays. Moreover, this decay can be dominant and can reach unity as 
shown in the right-pane of this figure. Despite this, such a decay may not give Higgs signals with sufficient statistical significance 
at the LHC (as discussed in previous literature and recalled here previously). Therefore, we are here well motivated to look further at the scope 
of direct production of $a_1$ state in single mode at the LHC, through $gg\to b\bar b a_1$, over overlapping regions of NMSSM parameter space,
which we are going to do next.

\section{Signal-to-Background Analysis}
\label{sect:S2B}

We perform here a partonic signal-to-background ($S/B$) analysis, based on CalcHEP, in the two channels $\gamma\gamma$ and
$\tau^+\tau^-$ separately in the two forthcoming subsections. We assume $\sqrt s=14$ TeV throughout for the LHC energy and we will benchmark
event rates on the basis of 300 inverse femtobarn of accumulated luminosity.

From the output of NMSSMTools, we have chosen some points which have large (yet not maximal) event rates as illustrative examples to test the 
possibility of detecting $a_1$ with different masses at the LHC. They are located in large regions of parameter space with high density.
These illustrative points are as follows:

\begin{itemize}
\item For $m_{a_1}$=9.8 GeV:\\ 
$\lambda$ = 0.22341, $\kappa$ = 0.41849, tan$\beta$ = 53.82, $\mu$ = 228.94,
 $A_\lambda$ = $-415.57$ and $A_\kappa$ = $-6.18$.
\item For $m_{a_1}$=20 GeV:\\
 $\lambda$ = 0.07595, $\kappa$ = 0.11544, tan$\beta$ = 51.51, $\mu$ = 377.44,
 $A_\lambda$ = $-579.64$ and $A_\kappa$ = $-3.53$. 
\item For $m_{a_1}$=31 GeV:\\
$\lambda$ = 0.10861, $\kappa$ = 0.46542, tan$\beta$ = 48.06, $\mu$ = 222.99,
 $A_\lambda$ = $-952.6$ and $A_\kappa$ = $-7.21$. 
\item For $m_{a_1}$=46 GeV:\\ 
$\lambda$ = 0.14088, $\kappa$ = 0.25219, tan$\beta$ = 50.56, $\mu$ = 317.08,
 $A_\lambda$ = -$569.61$ and $A_\kappa$ = $-8.61$. 
\item For $m_{a_1}$=60.5 GeV:\\
 $\lambda$ = 0.17411, $\kappa$ = 0.47848, tan$\beta$ = 52.39, $\mu$ = 169.83,
 $A_\lambda$ = $-455.85$ and $A_\kappa$ = $-9.03$.
\item For $m_{a_1}$=81 GeV:\\
 $\lambda$ = 0.10713, $\kappa$ = 0.13395, tan$\beta$ = 44.72, $\mu$ = 331.43,
 $A_\lambda$ = $-418.13$ and $A_\kappa$ = $-9.71$.   
\end{itemize}

\subsection{The $\gamma\gamma$ channel}
Altogether, the fact that production cross section and BR decay are maximised each in different regions of the NMSSM parameter space
makes it extremely difficult to obtain detectable event rates in this case. In fact, the signal yields for this channel are
not only small in general, but also overwhelmed by the irreducible background. This is made explicit in Figs. 5-6,
which show how the signal, despite yielding sizable peaks in the di-photon invariant mass $m_{\gamma\gamma}$ (for example, for
all $m_{a_1}$ masses considered -- between $\approx10$ and $\approx46$ GeV -- one can obtain some ${\cal O}(10)$ events every 300 fb$^{-1}$, see top-left
plot of each figure), is completely spoilt  by the irreducible background, after using standard cuts\footnote{Hereafter, $\eta$
refers to the pseudorapidity and $\Delta R$ to the cone distance expressed in differences of pseudorapidity and azimuthal angle $\phi$:
i.e., $\Delta R=\sqrt{(\Delta\eta)^2+(\Delta\phi)^2}$. Further, the notation $P_T$ refers to the transverse 
momentum.}:

$$\Delta R(b,\bar b),\Delta R(b,\gamma),\Delta R(\bar b,\gamma), \Delta R(\gamma,\gamma)> 0.4,$$
$$\arrowvert\eta(b)\arrowvert,\arrowvert\eta(\bar b)\arrowvert,\arrowvert\eta(\gamma)\arrowvert <2.5,$$
\begin{equation}
P_{T}(b),P_{T}(\bar b)>20~{\rm{GeV}}, P_{T}(\gamma)>2~{\rm{GeV}}.\label{cuts:AA}
\end{equation}

In fact, for triggering purposes, at least one of the two photon transverse momenta ought to be increased to some 10 GeV or so,
which does not help to improve $S/B$, as can be appreciated in the top-right and bottom plots in Figs. 5-6,
which show the average transverse momentum of a photon as well as its minimum and maximum, respectively. 
For $a_1$ masses above 50 GeV or so the signal rates are (in general) too poor to even pass the observability threshold of 1 event
after 300 fb$^{-1}$. 

In summary, despite the uniqueness of this signal, with the potential of clearly highlighting a possible singlet nature of the lightest
CP-odd Higgs of the NMSSM, the latter is not extractable at the LHC.

\subsection{The $\tau^+\tau^-$ channel}

The situation is instead much rosier for the $\tau^+\tau^-$ channel. After implementing the following 
standard cuts\footnote{Here, for the sake of illustration,
we take the $\tau$'s to be on shell.}
$$\Delta R (b,\bar b),\Delta R(b,\tau^{+}), \Delta R(\bar b,\tau^{+}),\Delta R(b,\tau^{-}), \Delta R(\bar b,\tau^{-}), \Delta R(\tau^{+},\tau^{-})> 0.4$$
$$\arrowvert\eta(b)\arrowvert,\arrowvert\eta(\bar b)\arrowvert,\arrowvert\eta(\tau^{+})\arrowvert,\arrowvert\eta(\tau^{-})\arrowvert <2.5$$
\begin{equation}
P_{T}(b),P_{T}(\bar b)>20~{\rm{GeV}}, P_{T}(\tau^{+}),P_{T}(\tau^{-})>10~{\rm{GeV}},
\label{cuts:TT}
\end{equation}
we obtain the invariant masses of the $\tau^+\tau^-$ system depicted in Figs. 7-12 (see their left-hand sides), where
the signal clearly appears over the irreducible background due to $q\bar q,gg\to b\bar b \gamma,Z\to b\bar b \tau^+\tau^-$. Recalling
that for low Higgs masses, say below $M_Z$, in the $b\bar b\tau^+\tau^-$ channel, the dominant background is 
indeed the irreducible one, see, e.g., Ref.~\cite{bbHiggs-lowmass}, at least in the double- and single-leptonic channels,
it is clear that there exist substantial discovery potential of a very light CP-odd Higgs boson of the NMSSM at the LHC.

For the sake of completeness, we have also shown in Figs.~7-12 the top-antitop reducible background,
i.e., $q\bar q,gg\to t\bar t\to b\bar b W^+W^-\to b\bar b \tau^+\tau^- P_T^{\rm{miss}}$ (with no cuts on $P_T^{\rm{miss}}$ though,
which could always be enforced for its suppression),
making it clear that discovery of heavier $a_1$ states in $\tau^+\tau^-$ pairs, even in the the double- and single-leptonic channels,
becomes much more difficult, although not impossible (also owing to the long $Z$ tail in the irreducible background).
 The case of fully hadronic decays is further
contaminated by pure QCD backgrounds, the more so the smaller the $m_{a_1}$ values, so that we have not treated it here. Finally notice the
stability of the aforementioned signals against the possibility of stiffer thresholds in the selection of $\tau$ decays, as is made clear
by the $P_T(\tau)$ dependence of the $S$ and $B$ distributions in the right-hand side of  Figs. 7-12 (e.g., an
increase of even a factor of two in transverse momentum of the leptonic/hadronic $\tau$ decay products should not 
dramatically spoil the signal significances in the low $a_1$ mass region). 

Overall, the $\tau^+\tau^-$ signal yield in the low $a_1$ mass region is of order 3000 (for $m_{a_1}$ reaching 80 GeV or so)  to 30000
 (for $m_{a_1}$ starting at 10 GeV or so) signal events over a much smaller background, even assuming $\tau^+\tau^-$ resolutions of 10 GeV or so
(notice that the width of each histogram in the figures is of 1 GeV and the plots are in log scale).  

\section{Conclusions}
\label{sect:summa}
In short, we have proven that there exist some regions of the NMSSM parameter space where very light CP-odd Higgs states, with a mixed
singlet and doublet nature, could potentially be detected if $m_{a_1}\le M_Z$ in the $a_1\to\tau^+\tau^-$ mode if the CP-odd Higgs state is produced
in association with a $b\bar b$ pair. 
After a realistic $S/B$ analysis at parton level, we have in fact produced 
results showing that
the extraction of light mass $a_1\to\tau^+\tau$ resonances above both the irreducible and (dominant) reducible backgrounds should be feasible using standard
reconstruction techniques \cite{ATLAS-TDR,CMS-TDR}, at least in the double- and single-leptonic decay channels of the $\tau$'s. While more refined
analyses, incorporating $\tau$-decay, parton shower, hadronisation and detector effects, are needed in order to delineate the 
true discovery potential of the LHC
over the actual NMSSM parameter space, we believe that our results are a step in the right direction to prove the existence of a `More-to-gain
theorem' at the CERN collider for the NMSSM with respect to the MSSM, as $\tau^+\tau^-$ signals from such light Higgs bosons are not possible
in the latter scenario.
Finally, some of the parameter regions where the aforementioned signal can be detected overlap with those where 
$h_{1,2}\to a_1a_1$ decays could also be effective in extracting an $a_1$ signal, so that the process
discussed here also offers an alternative handle to establish a `No-lose theorem' for the NMSSM at the LHC.

Further notice that we have explored here the two regimes $2m_\tau < m_{a_1} < 2m_b$ and $2m_b < m_{a_1}$.
The former is where the $a_1\to \tau^+\tau^-$ decay rate dominates (these are the points in the `red island' to the outermost 
right-hand side of the  middle-right plot of Fig.~3), for which BR$(a_1\to \tau^+\tau^-)\ge0.9$, as the $b\bar b$ decay channel is closed.
(Our mass point $m_{a_1}=9.8$ GeV was representative of this situation). The latter is where  the $a_1\to b\bar b$ decay rate dominates 
(the corresponding points are all the remaining ones of Fig.~3), since here BR$(a_1\to \tau^+\tau^-)\le0.1$ because the $b\bar b$ decay channel is 
open. (Our mass points $m_{a_1}=20$, 31, 46, 60.5 and 81 GeV were representative of this situation.) 

Unfortunately, a similar analysis in the $a_1\to\gamma\gamma$ channel has showed that the LHC discovery potential is hindered by an
overwhelming irreducible background, despite the fact that the BR$(a_1\to\gamma\gamma)$ can be equal to unity in some regions of the NMSSM
parameter space, a peculiarity of the NMSSM with respect to the MSSM.

\section*{Acknowledgments}
We acknowledge useful email exchanges with Cyril Hugonie and Ulrich Ellwanger. We are also very grateful to Lorenzo Basso for technical assistance
with some of the plots and to Alexander Belyaev for his help with CalcHEP.
This work is 
supported in part by the NExT Institute. M. M. A. acknowledges
a scholarship granted to him by Taibah University (Saudi Arabia).

\newpage

\begin{figure}[H]
 
 \centering\begin{tabular}{cc}
 \includegraphics[scale=0.75]{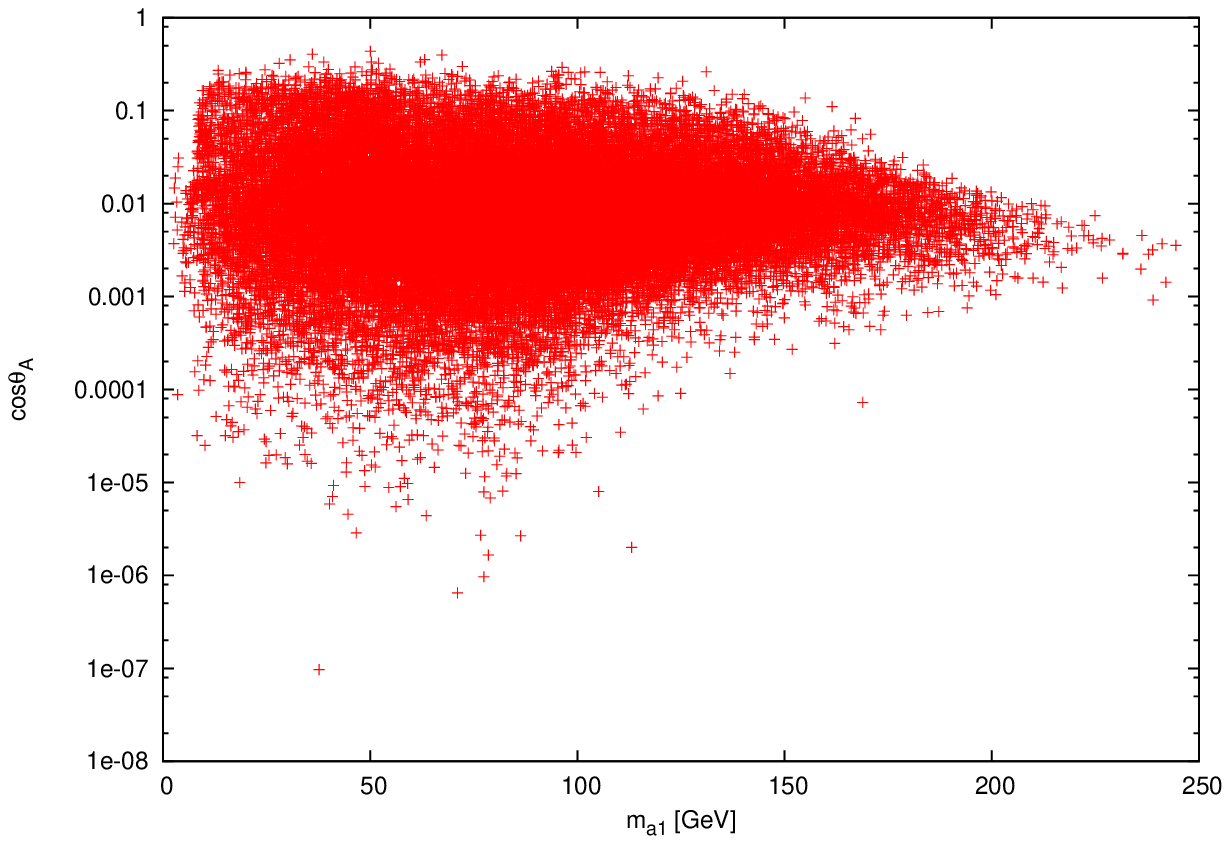}\\
\includegraphics[scale=0.75]{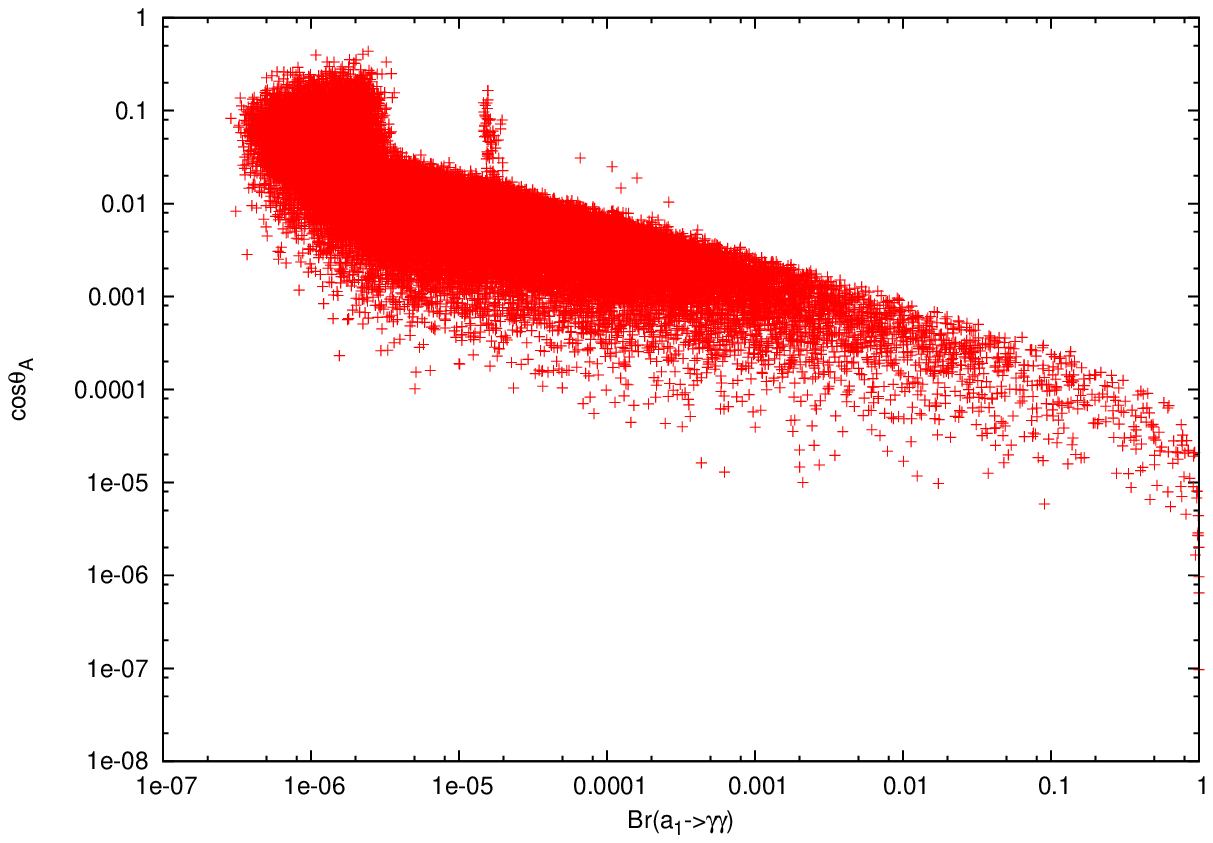}
 \end{tabular}
\label{fig:costheta}
\caption{The lightest CP-odd Higgs mass $m_{a_1}$ and the BR$(a_1\to \gamma\gamma)$ plotted against the mixing angle
in the CP-odd sector, $\cos\theta_A$. }
\end{figure}

\begin{figure}
 \centering\begin{tabular}{cc}
 \includegraphics[scale=0.15]{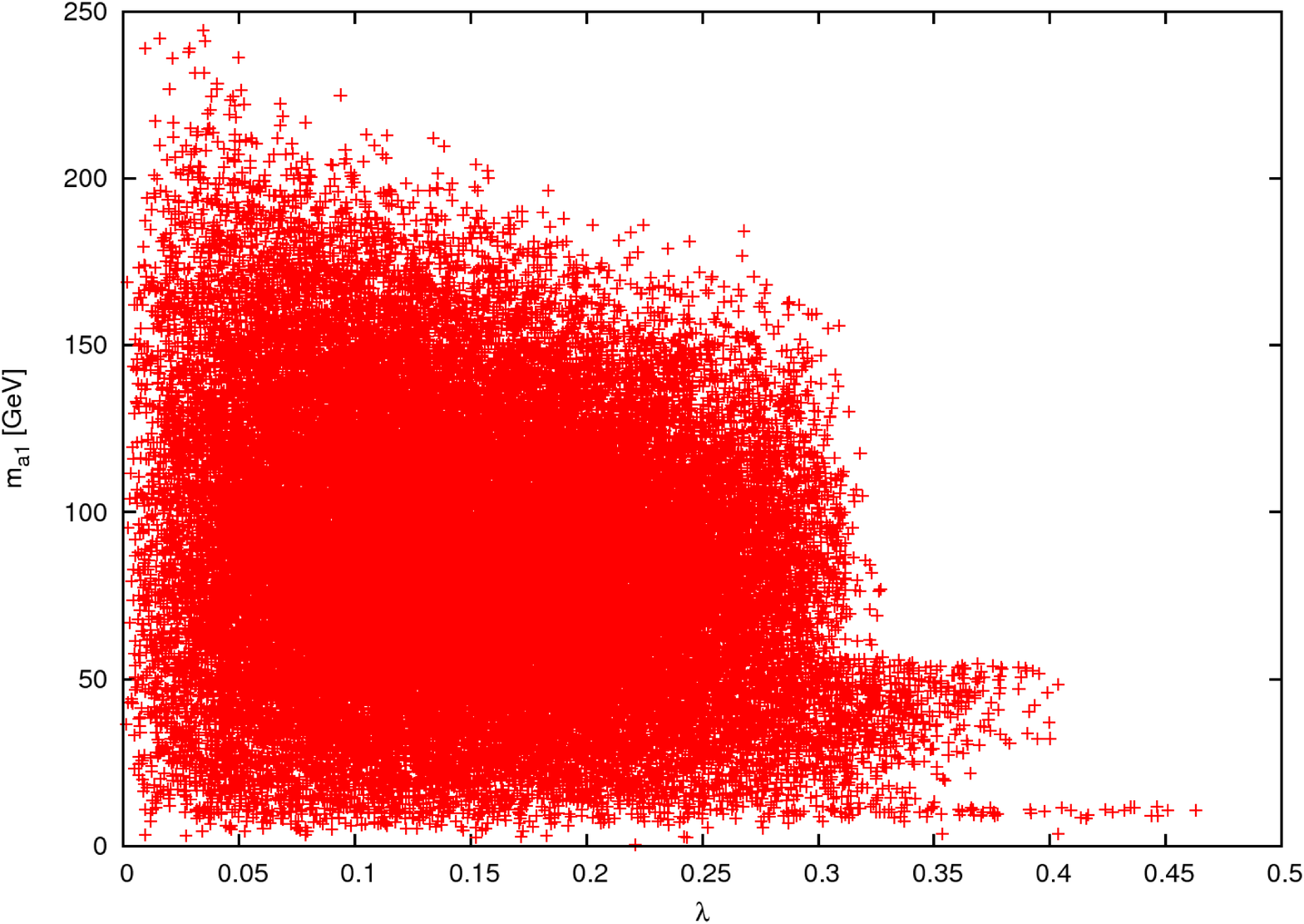}
&\includegraphics[scale=0.15]{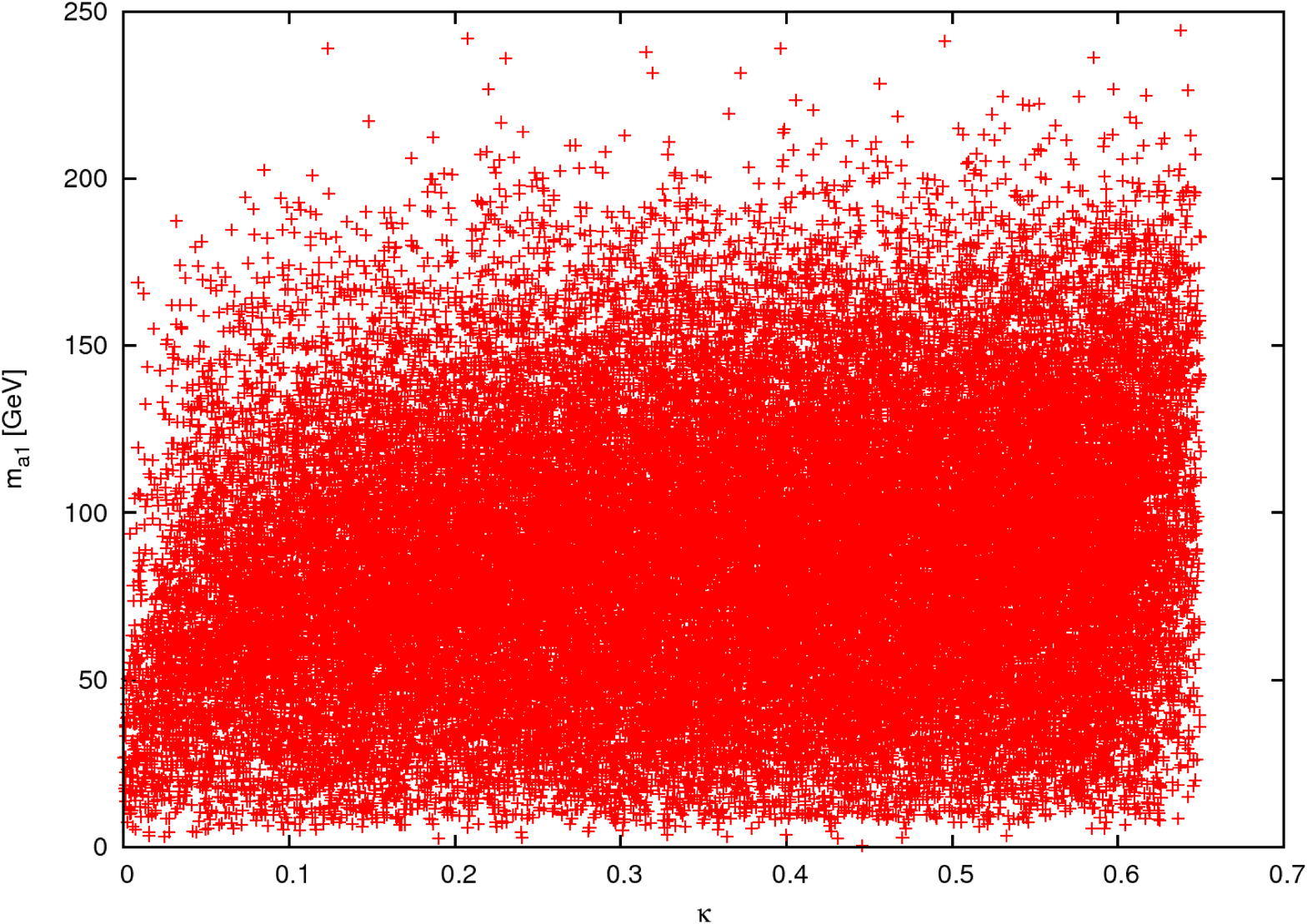}\\
  \includegraphics[scale=0.15]{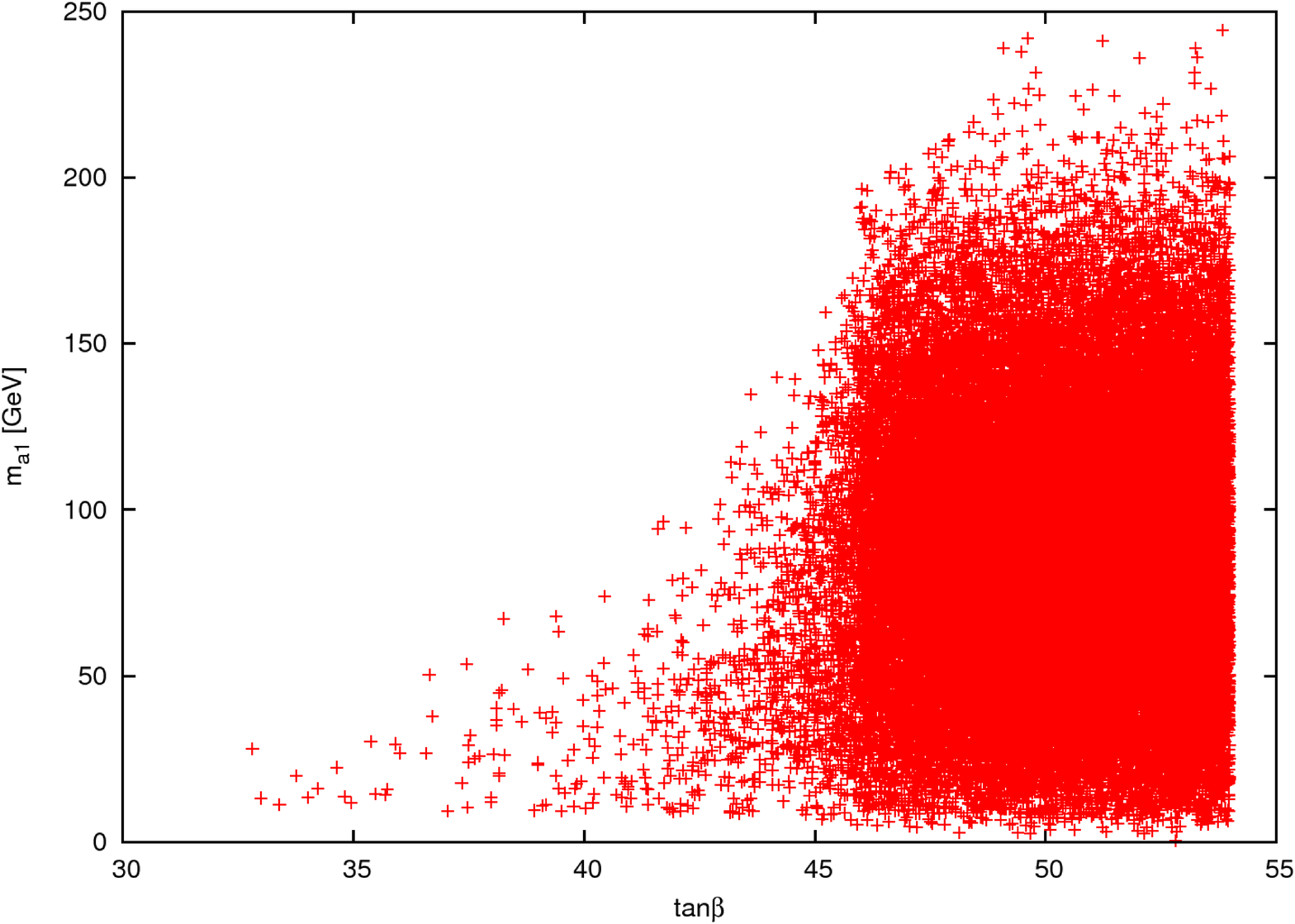}
&\includegraphics[scale=0.15]{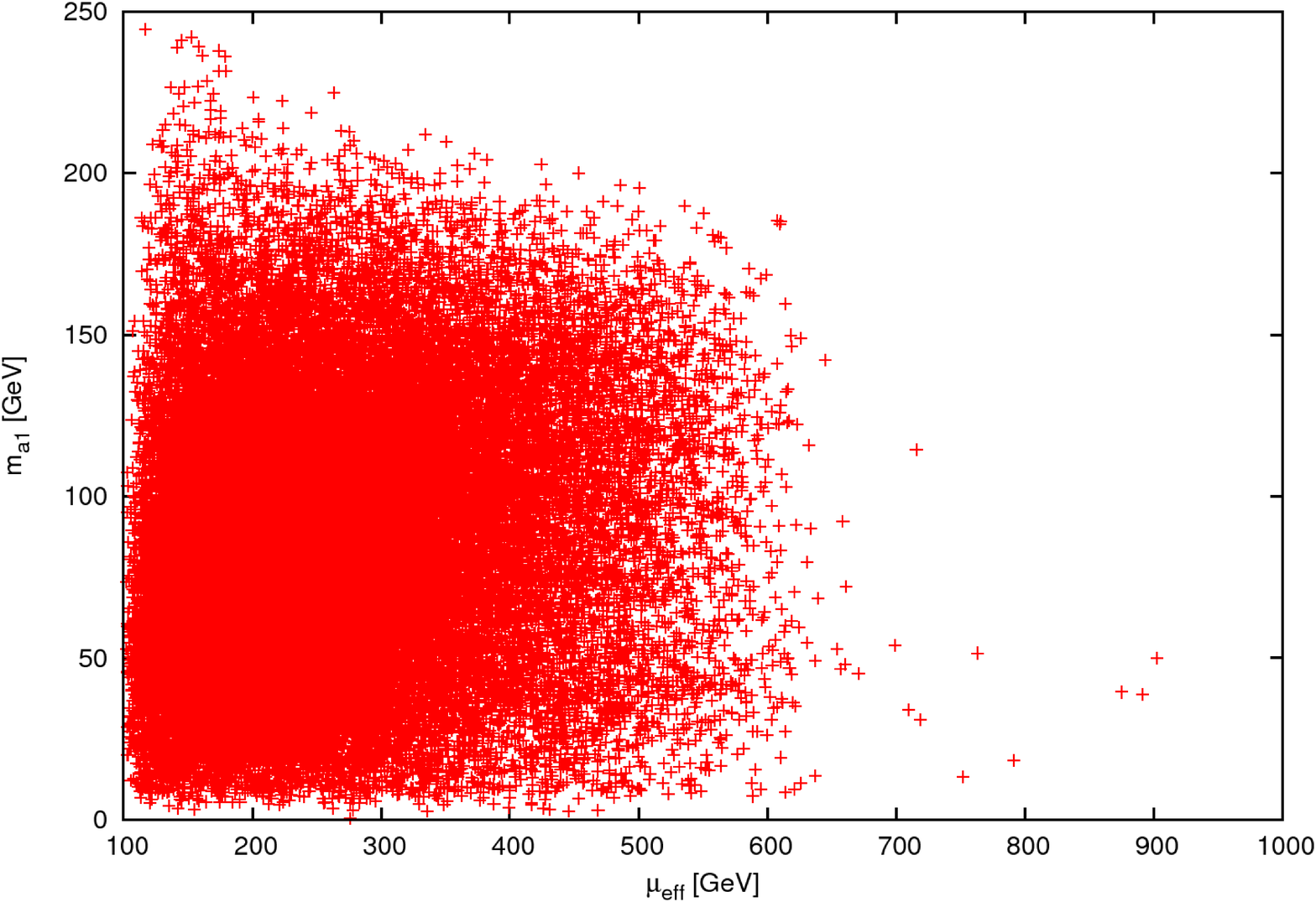}\\
  \includegraphics[scale=0.15]{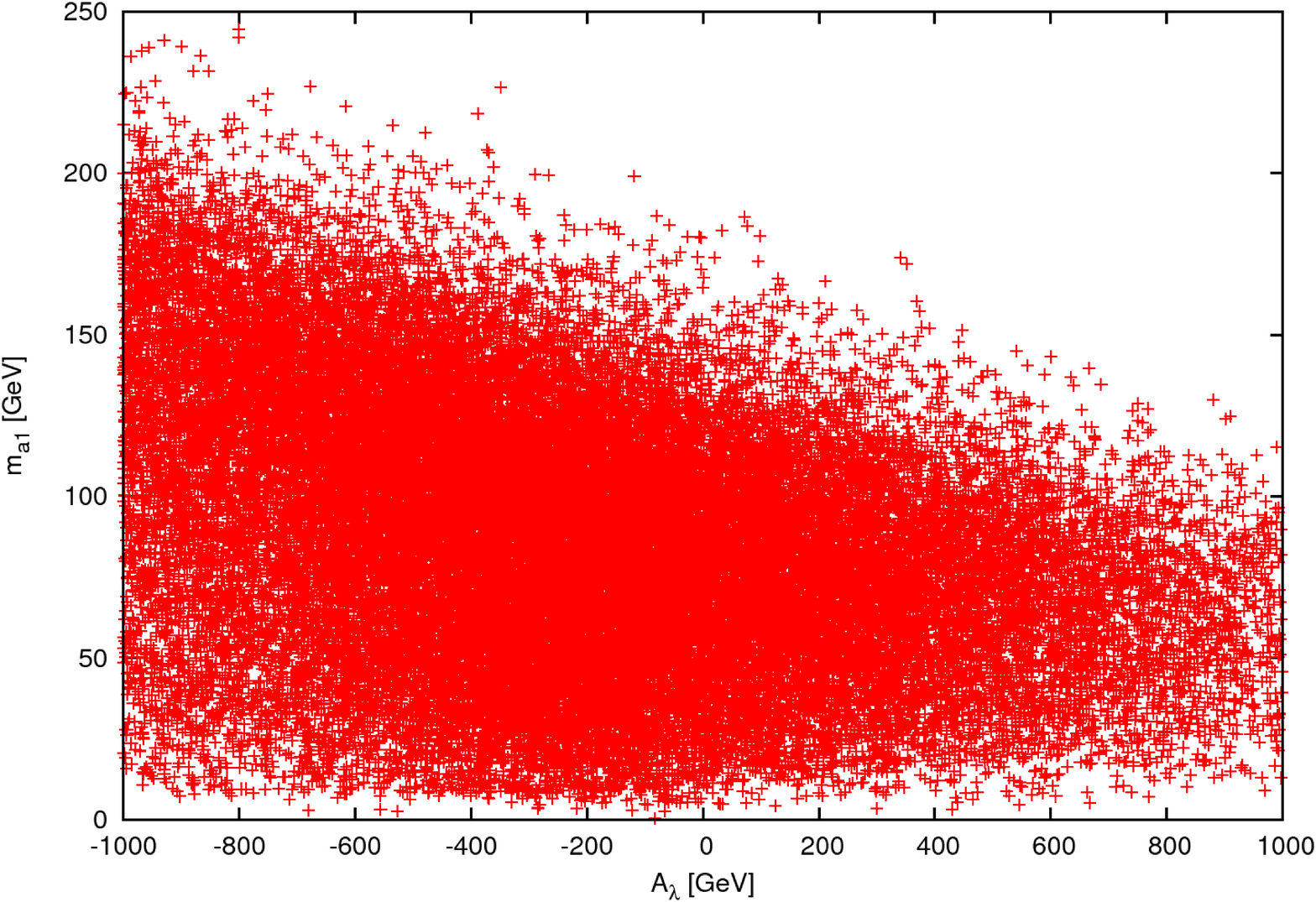}
&\includegraphics[scale=0.15]{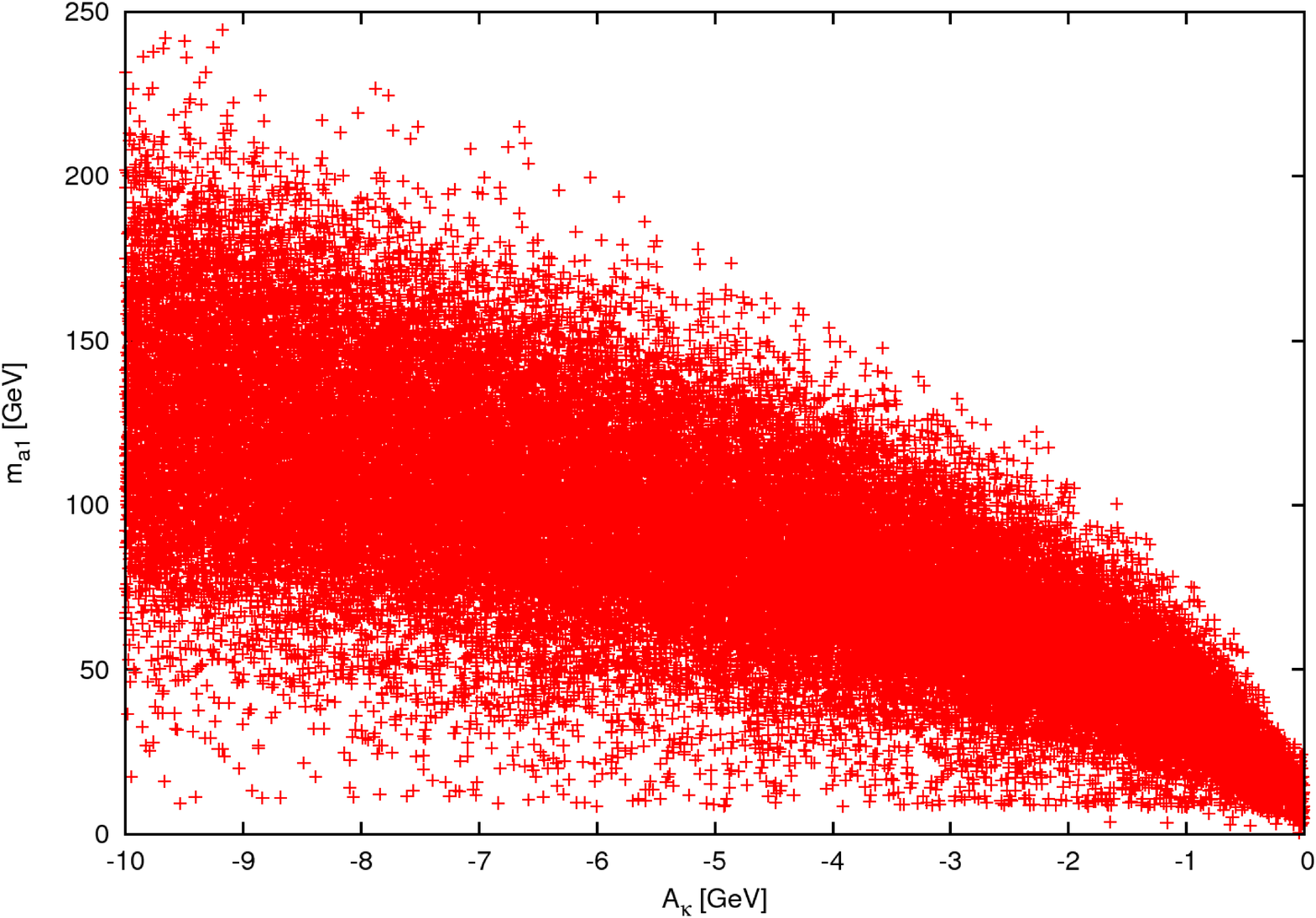}\\
 \includegraphics[scale=0.15]{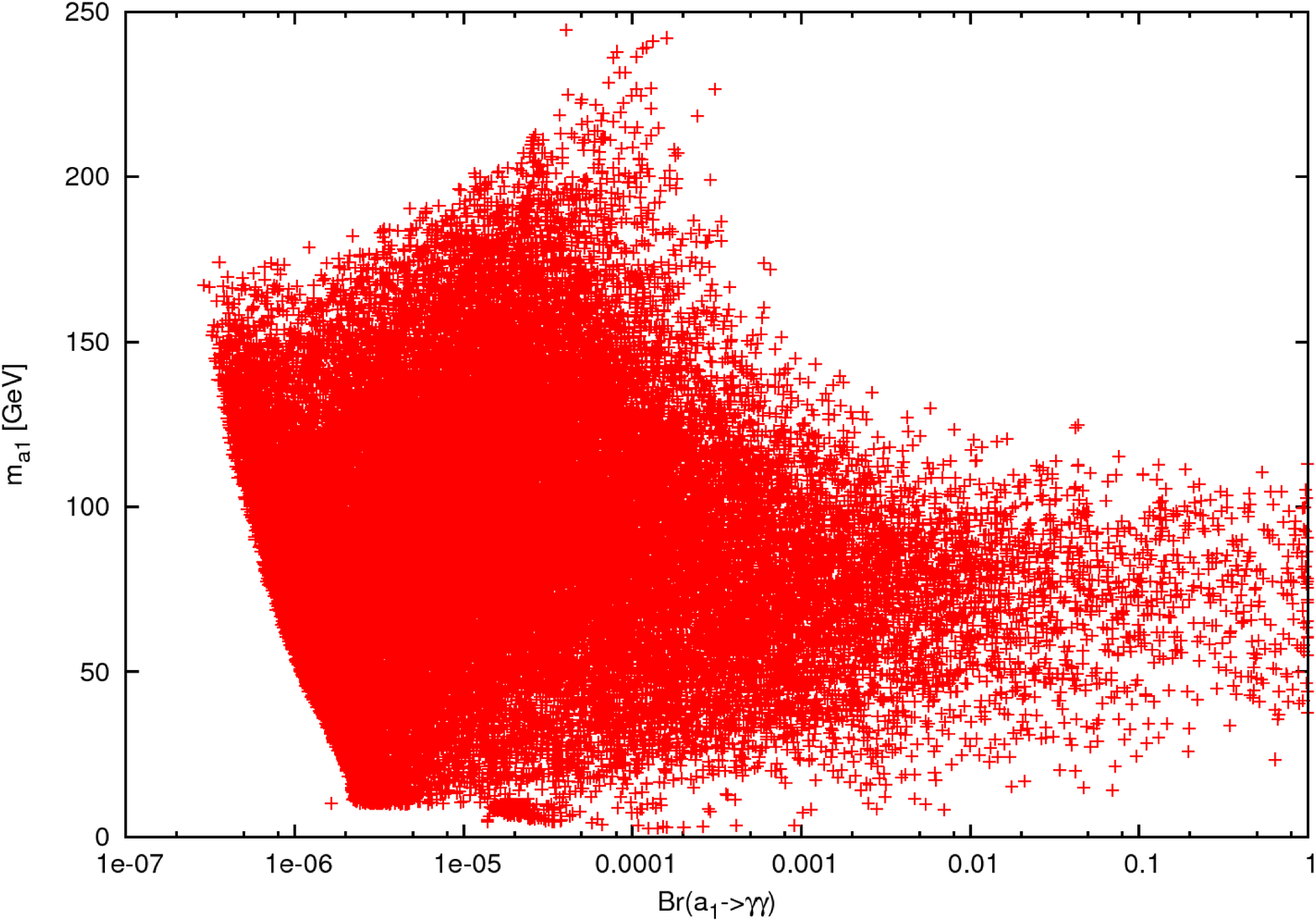}
 &\includegraphics[scale=0.15]{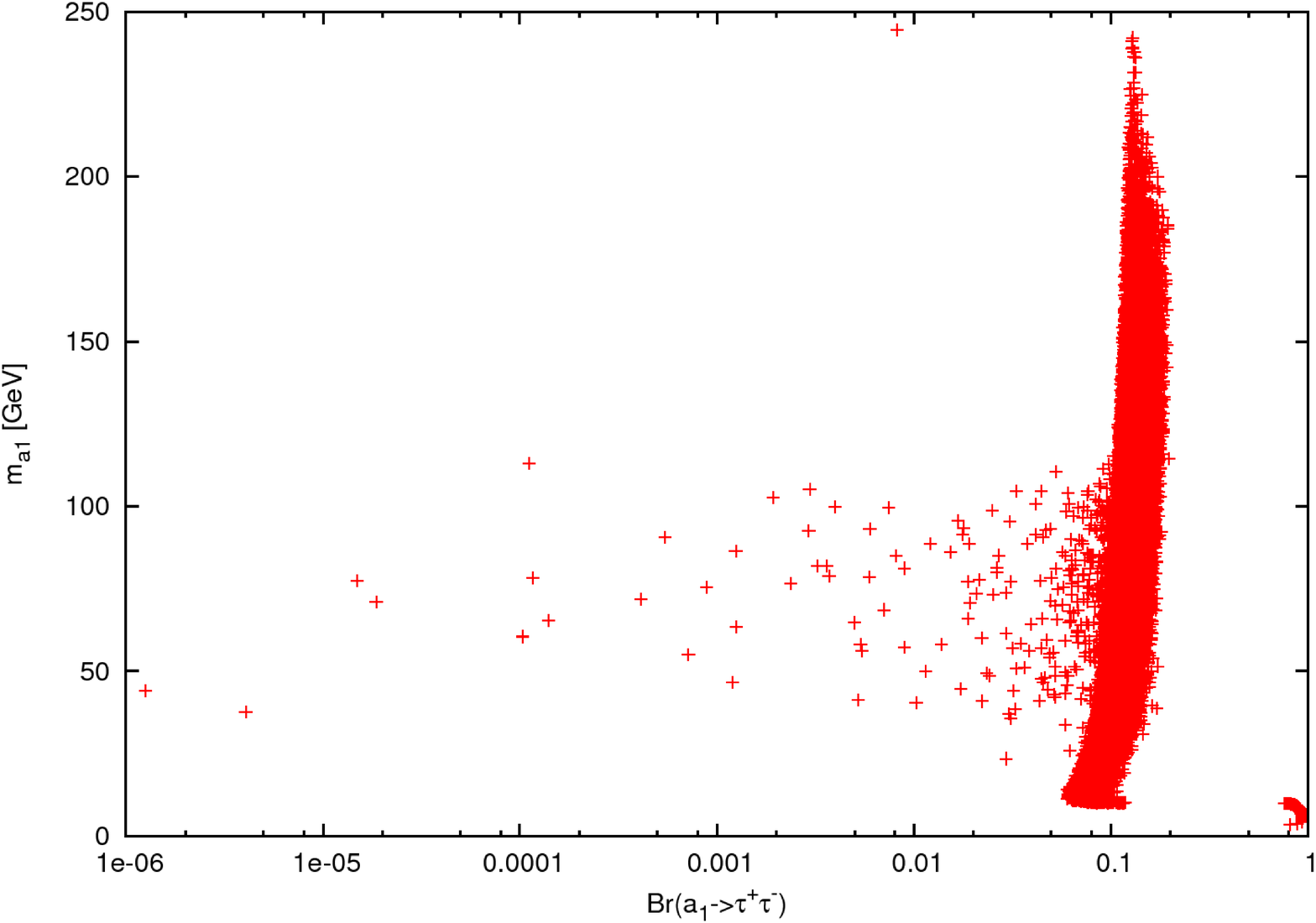}
 \end{tabular}
\label{fig:mass-scanAA}
\caption{The CP-odd Higgs mass $m_{a_1}$ as a function of $\lambda$, $\kappa$, $\tan\beta$, $\mu_{\rm eff}$,
 $A_\lambda$, $A_\kappa$, BR$(a_1\to \gamma\gamma)$ and BR$(a_1\to \tau^{+}\tau^{-})$.}
\end{figure}

\begin{figure} 
 \centering\begin{tabular}{cc}
\includegraphics[scale=0.15]{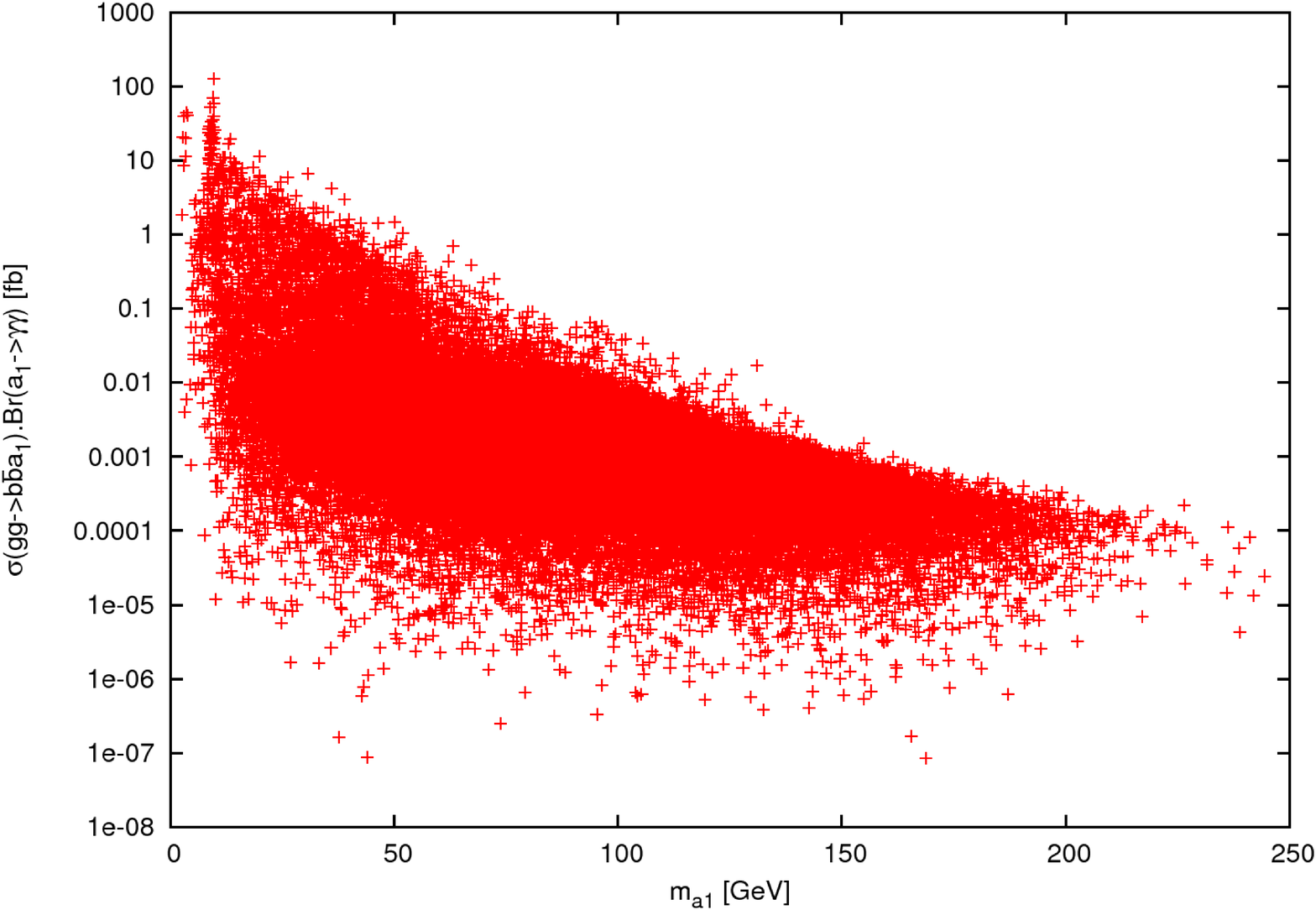}
&\includegraphics[scale=0.15]{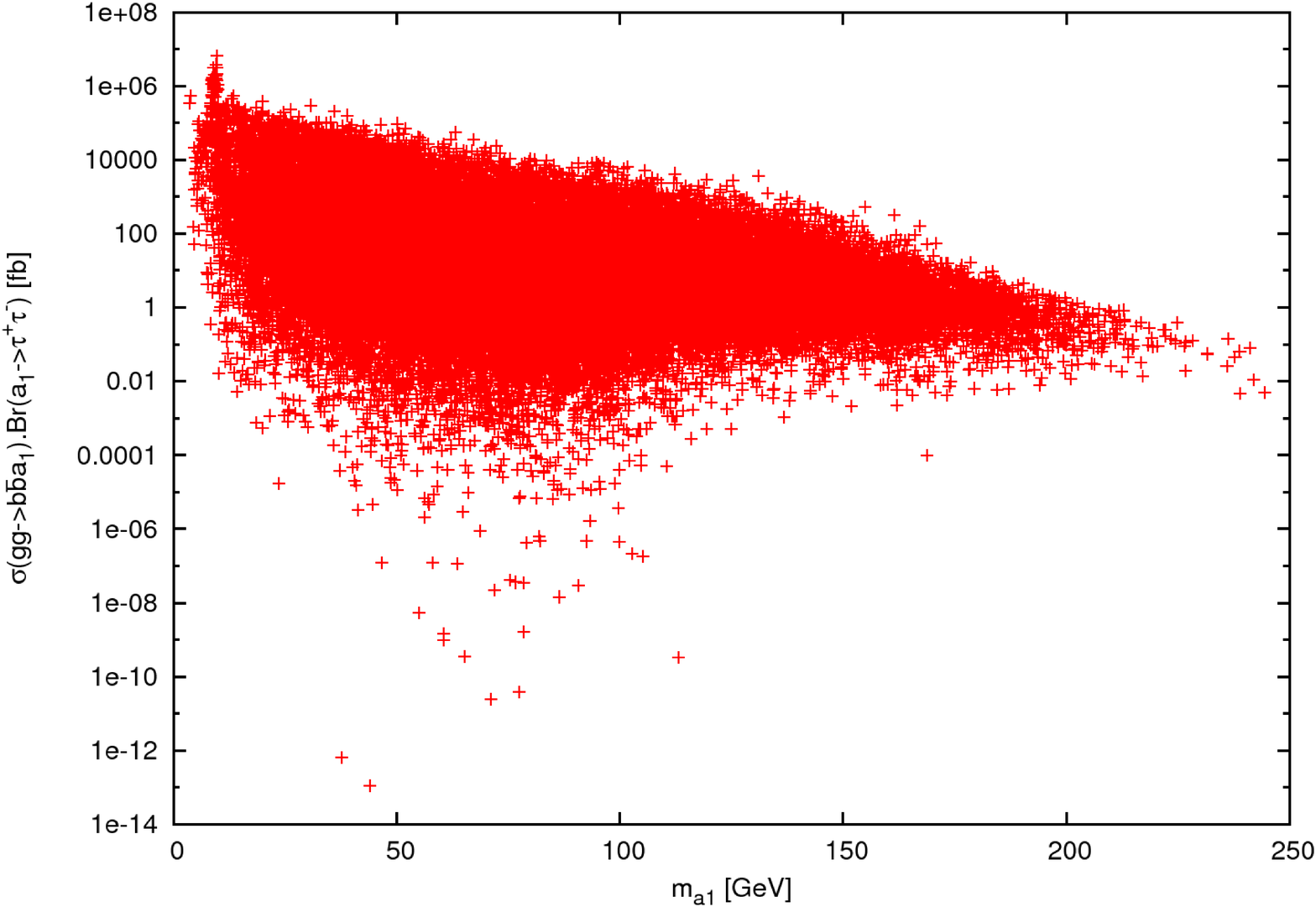}\\
\includegraphics[scale=0.15]{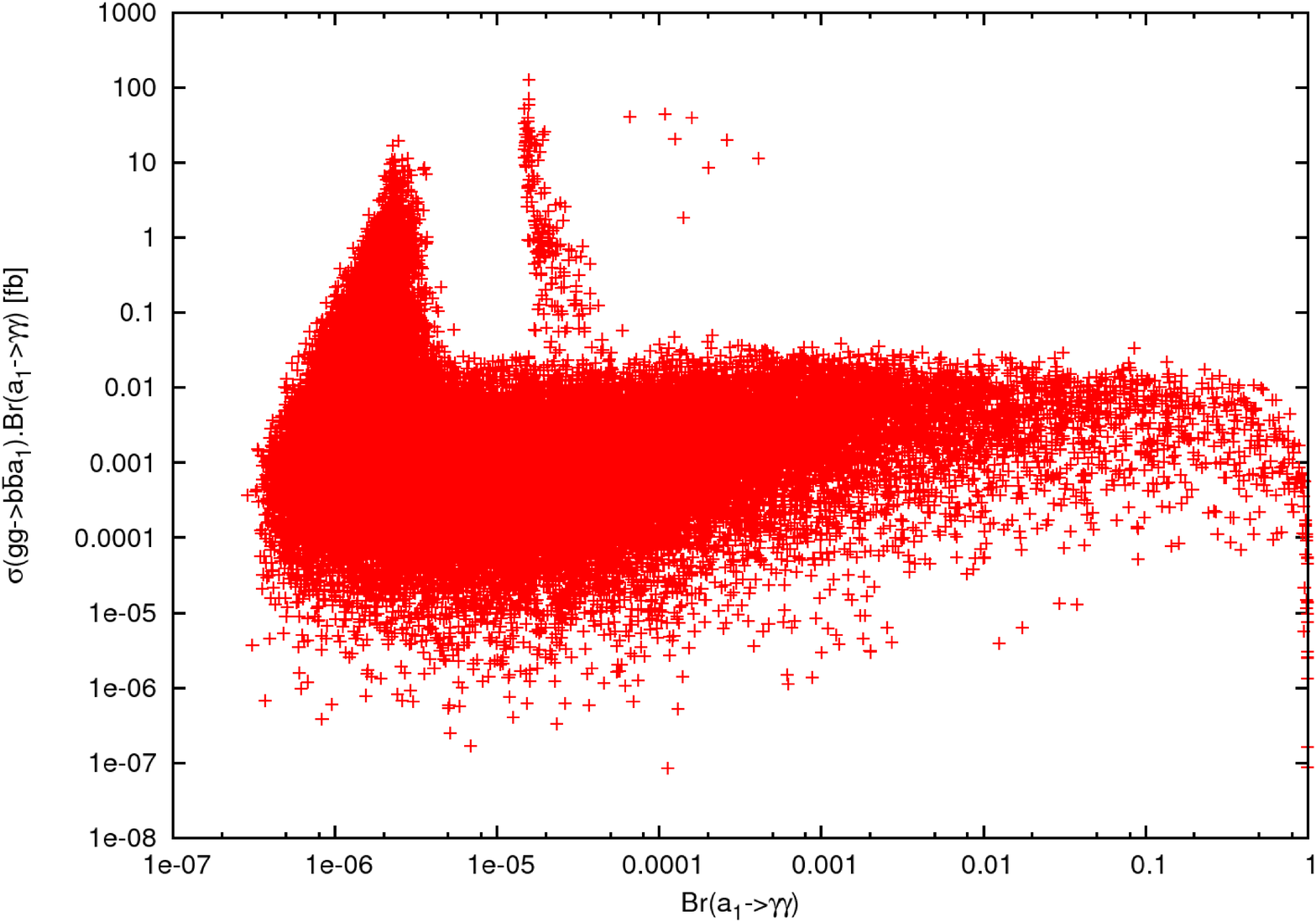}
&\includegraphics[scale=0.15]{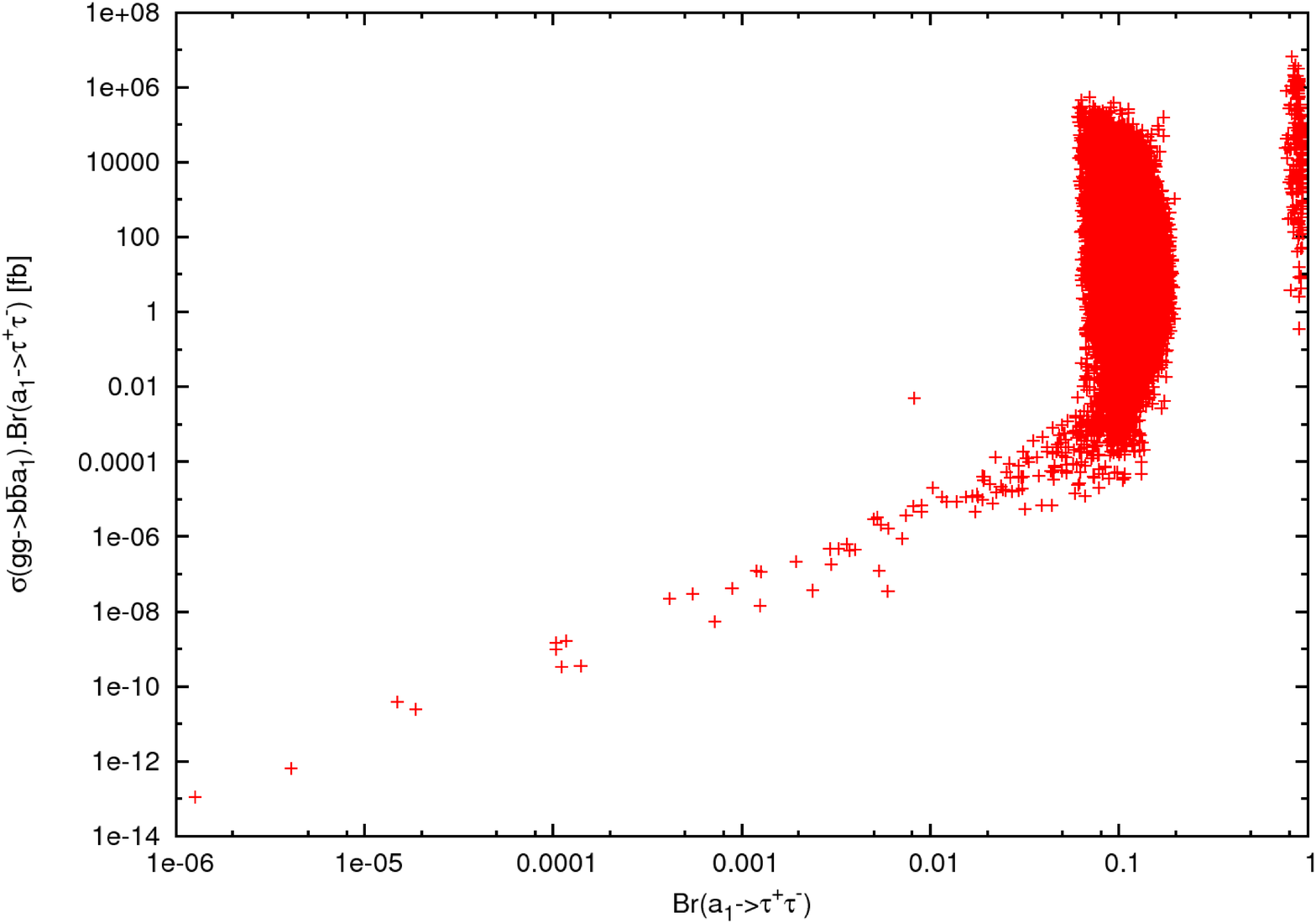}\\
\includegraphics[scale=0.15]{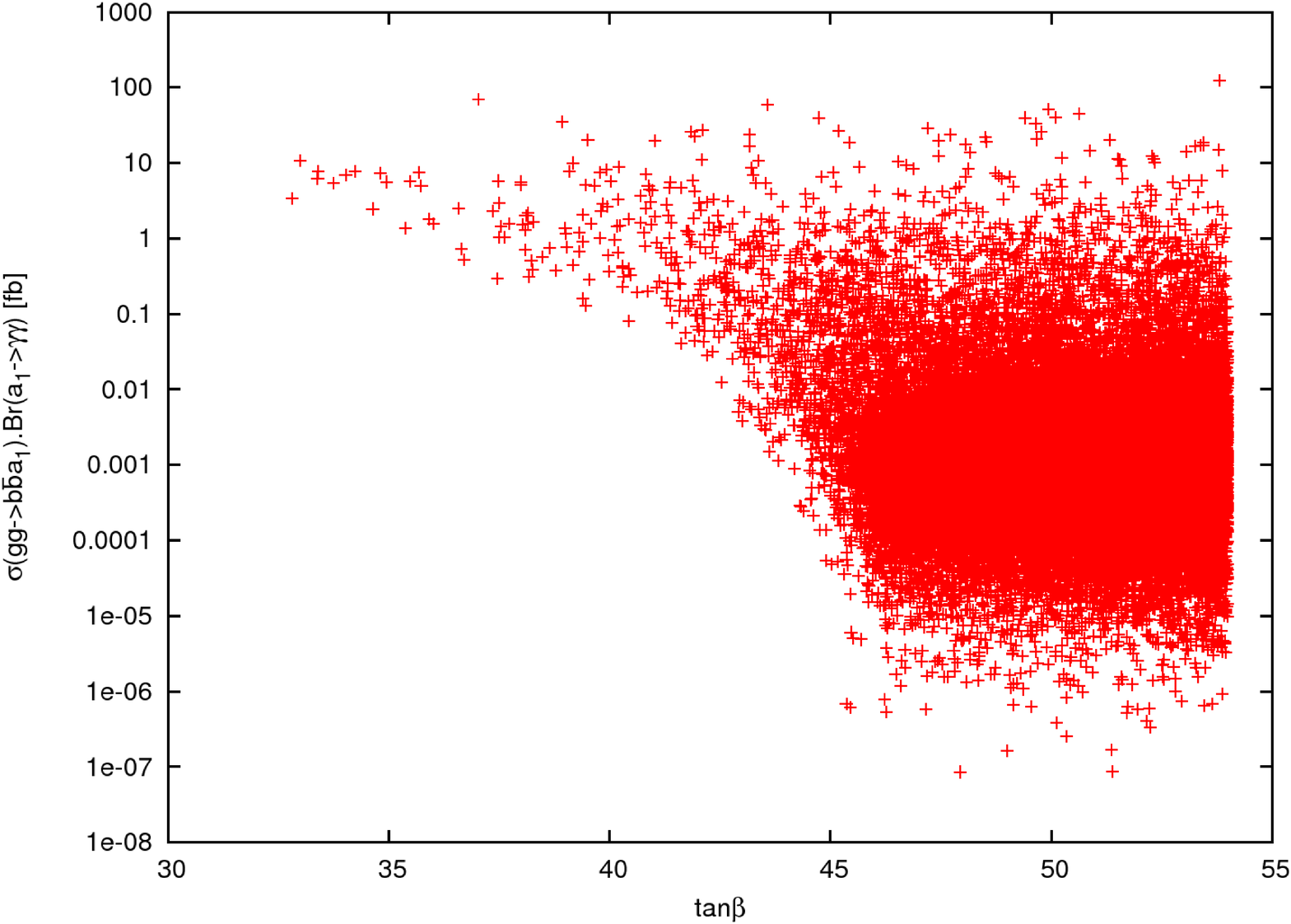}
&\includegraphics[scale=0.15]{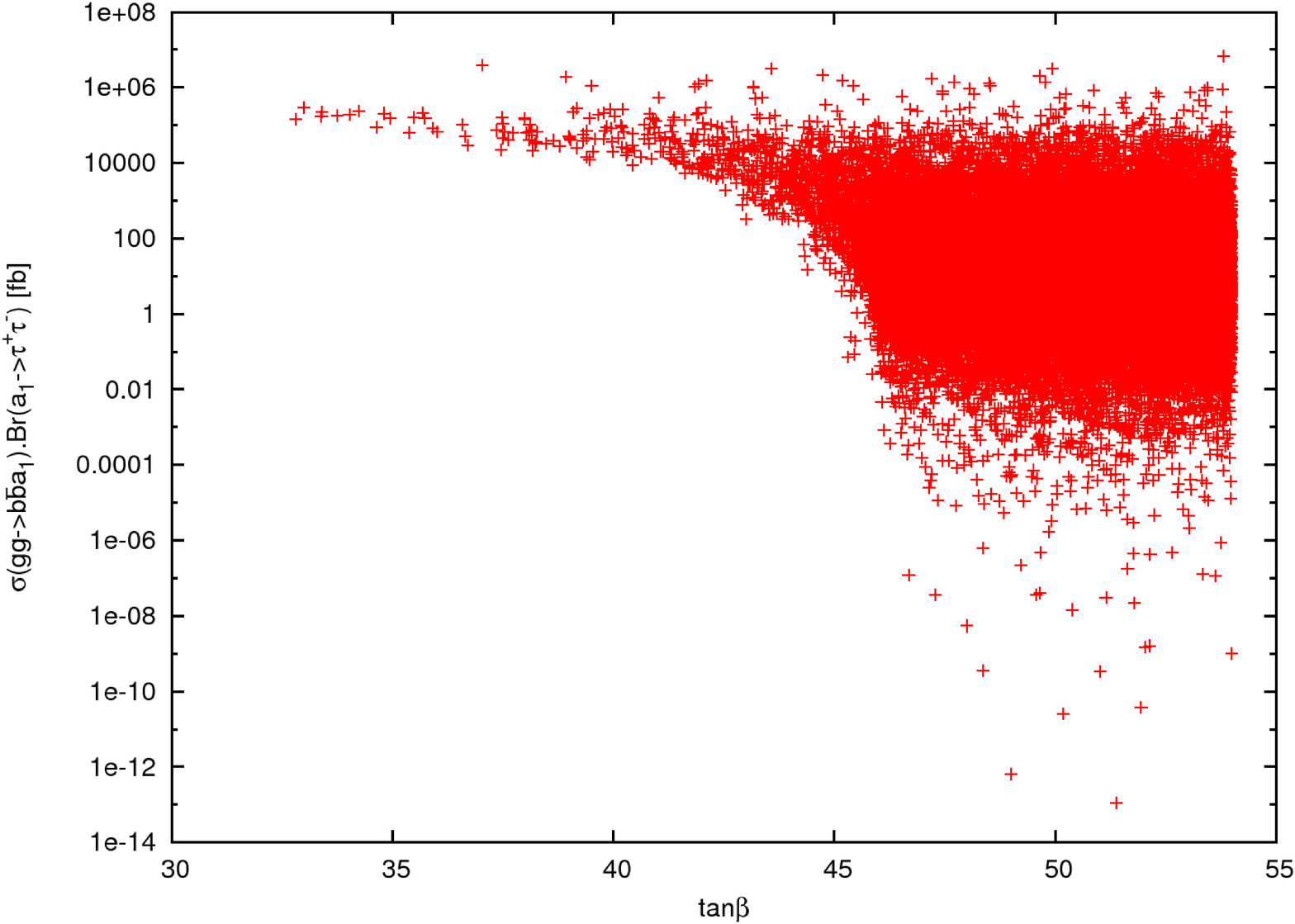}
 \end{tabular}
\label{fig:sigma-scanAA}
\caption{The rates for $\sigma(gg\to b\bar b {a_1})~{\rm BR}(a_1\to \gamma\gamma)$ (left) and
$\sigma(gg\to b\bar b {a_1})~{\rm BR}(a_1\to \tau^+\tau^-)$ (right) as functions of $m_{a_1}$, BR of the corresponding channel and
$\tan\beta$. }
\end{figure}

\begin{figure} 
 \centering\begin{tabular}{cc}
\includegraphics[scale=0.65]{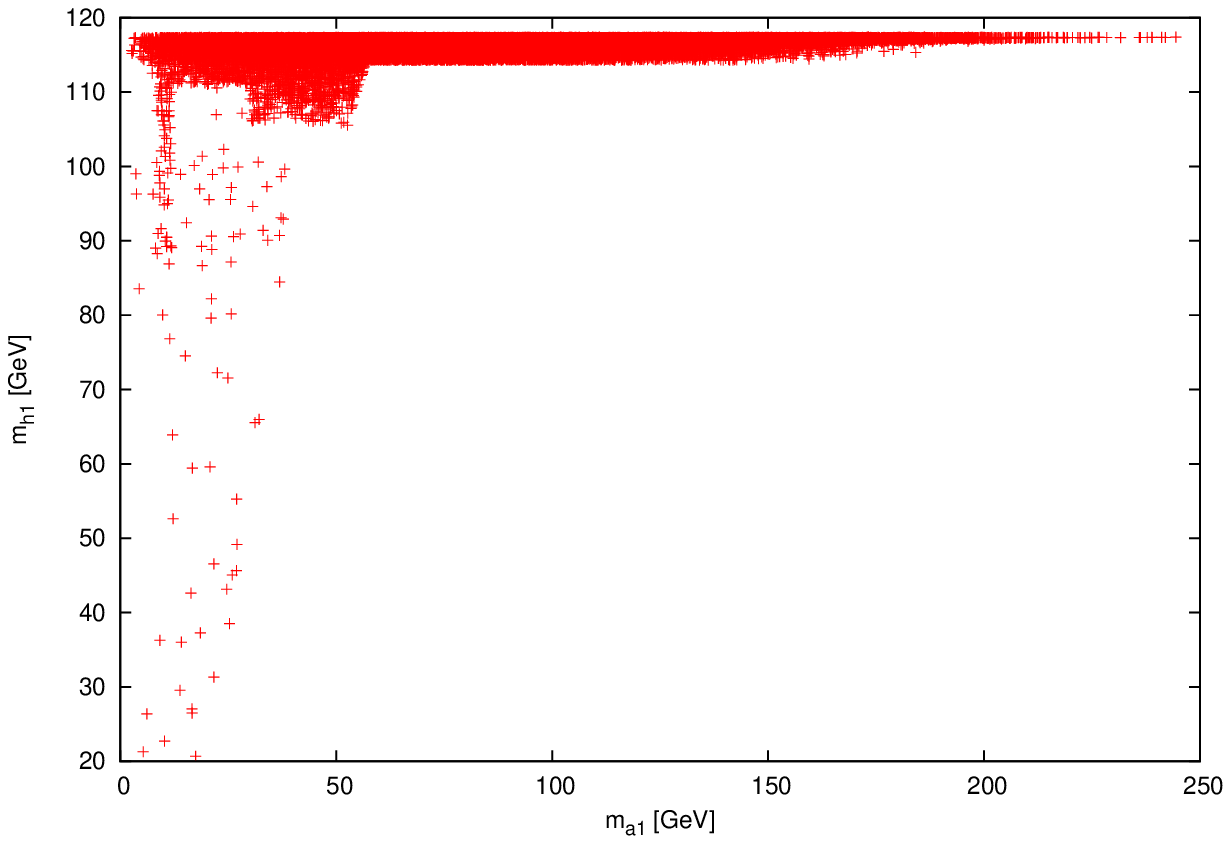}
&\includegraphics[scale=0.65]{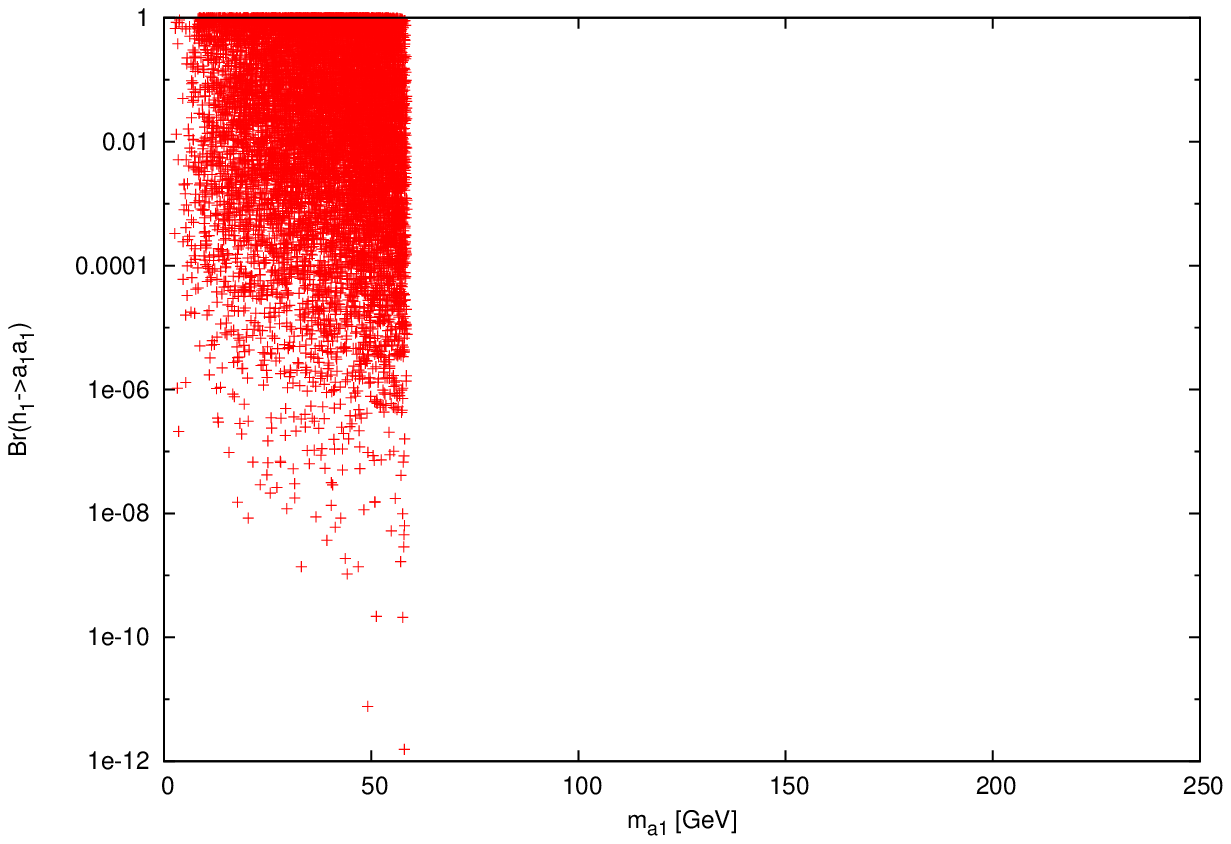}\\
 \end{tabular}
\label{fig:mh1-ma1}
\caption{The lightest CP-odd Higgs mass $m_{a_1}$ plotted against the lightest CP-even Higgs mass $m_{h_1}$ and BR$(h_1\to a_1a_1)$. }
\end{figure}

\newpage
\begin{figure}
 \centering\begin{tabular}{cc}
 \includegraphics[scale=0.75]{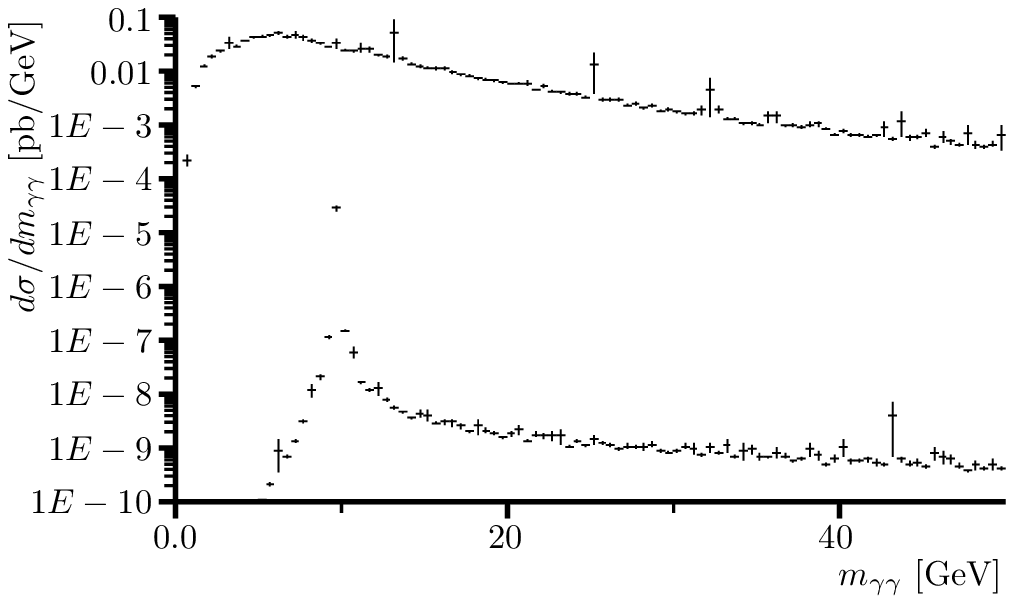}
&\includegraphics[scale=0.75]{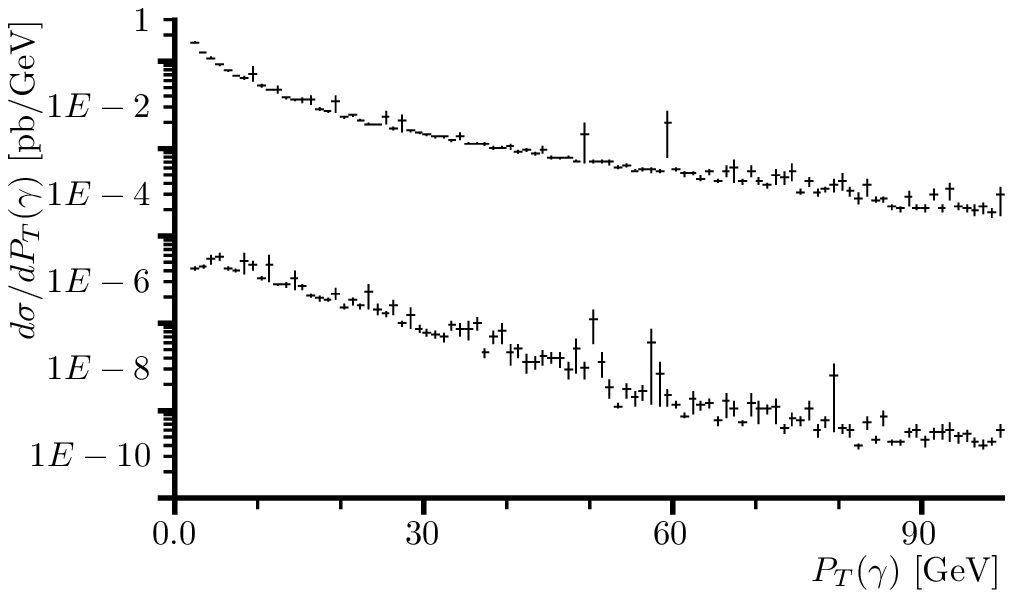}\\
   \includegraphics[scale=0.75]{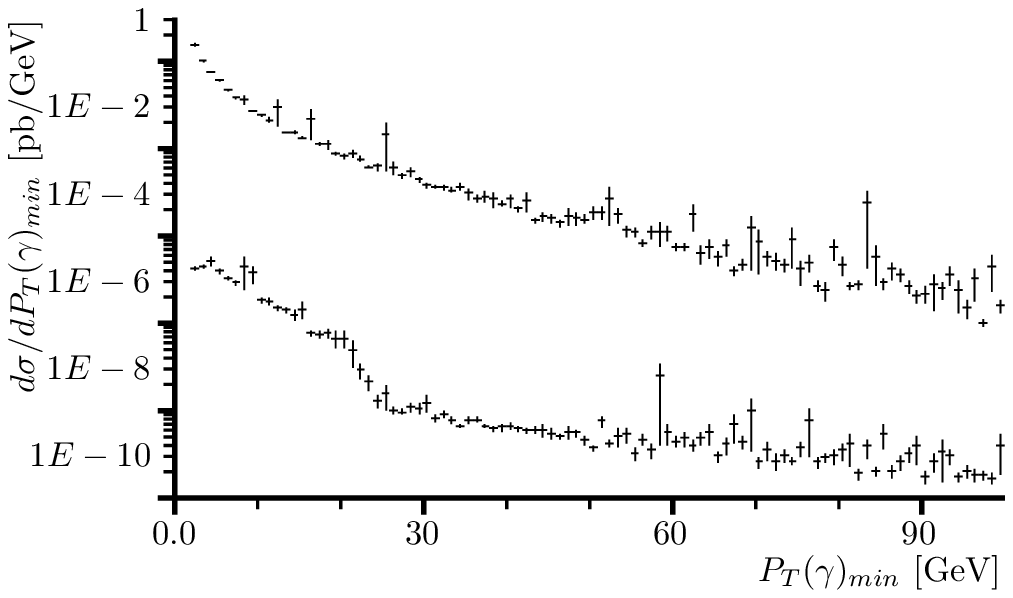}
&\includegraphics[scale=0.75]{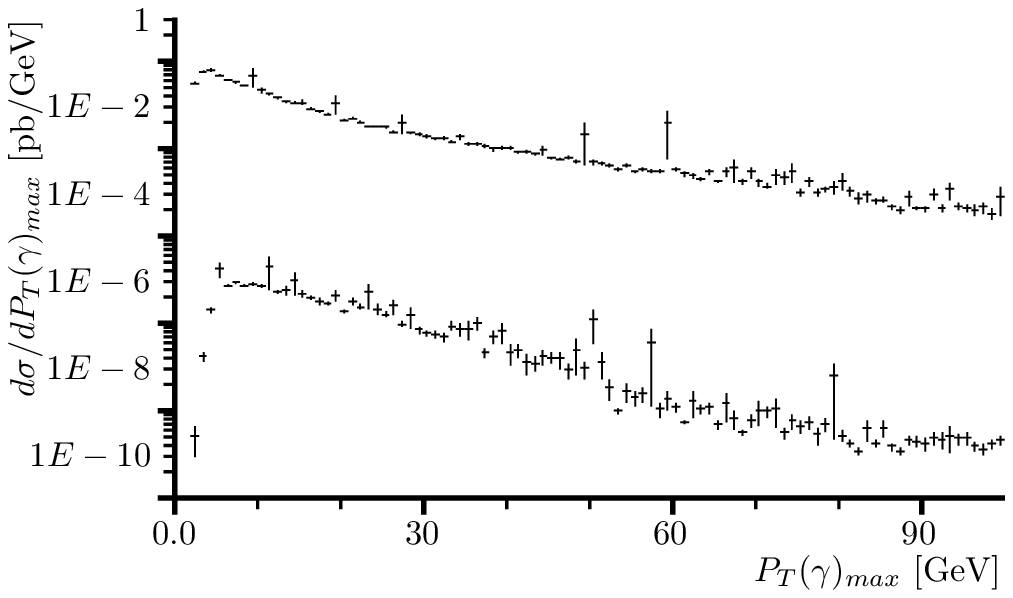}

 \end{tabular}  
\label{fig:AA1}
\caption{The differential cross section for $m_{a_1}$ = 9.8 GeV as a function of the invariant mass 
$m_{\gamma\gamma}$, $P_{T}(\gamma)$, $P_{T}(\gamma)^{\rm{min}}$ and $P_{T}(\gamma)^{\rm{max}}$ for the signal 
(bottom distribution) only and for the signal and the background together 
(top distribution), after the cuts in (\ref{cuts:AA}).} 
\end{figure}

\begin{figure}
 \centering\begin{tabular}{cc}
 \includegraphics[scale=0.75]{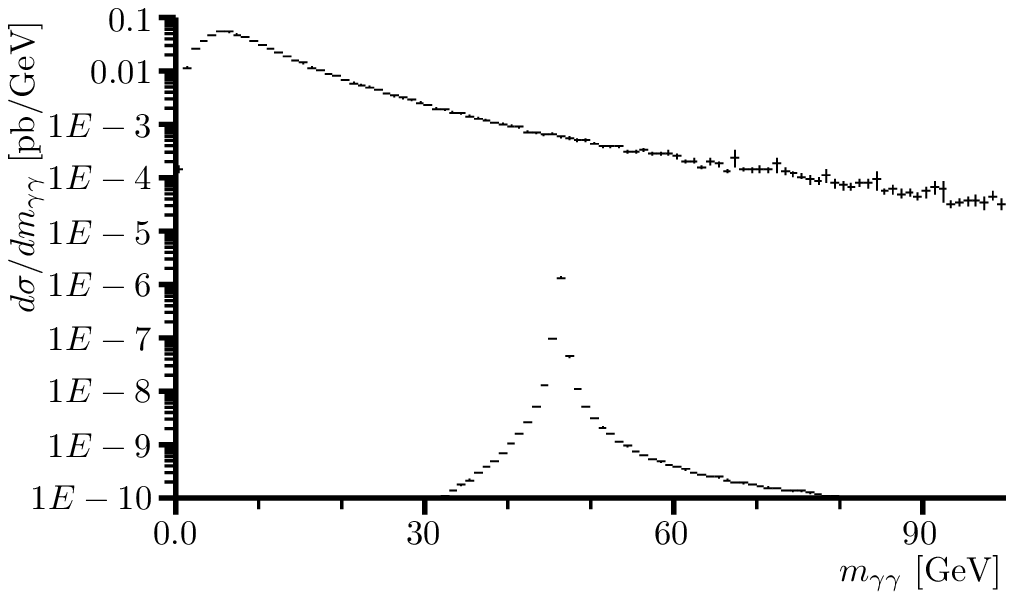}
&\includegraphics[scale=0.75]{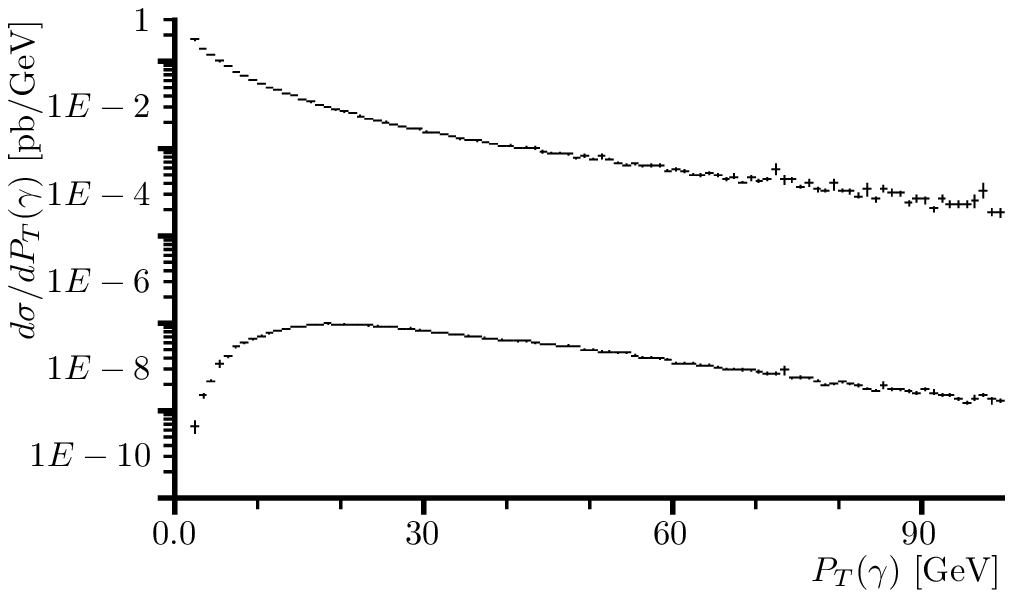}\\
   \includegraphics[scale=0.75]{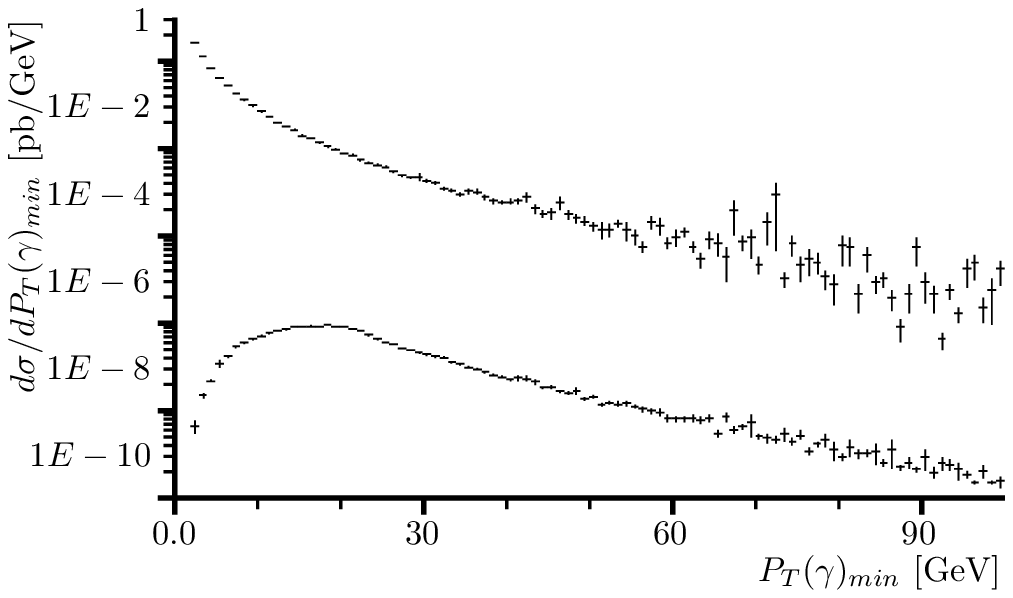}
&\includegraphics[scale=0.75]{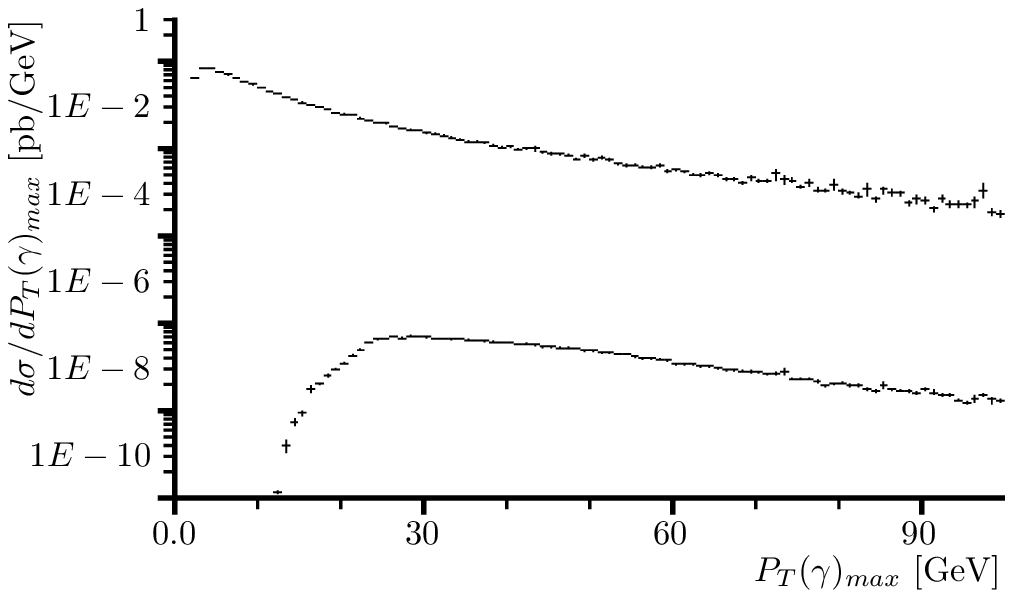}

 \end{tabular}
\label{fig:AA4}
\caption{The differential cross section in the $\gamma\gamma$ channel for $m_{a_1}$ = 46 GeV as a function of
 the invariant mass $m_{\gamma\gamma}$, $P_{T}(\gamma)$, $P_{T}(\gamma)^{\rm{min}}$ and $P_{T}(\gamma)^{\rm{max}}$
 for the signal (bottom distribution) only and for the signal and the background 
together (top distribution), after the cuts in (\ref{cuts:AA}). }
\end{figure}

\begin{figure}
 \centering\begin{tabular}{cc}

 \includegraphics[scale=0.75]{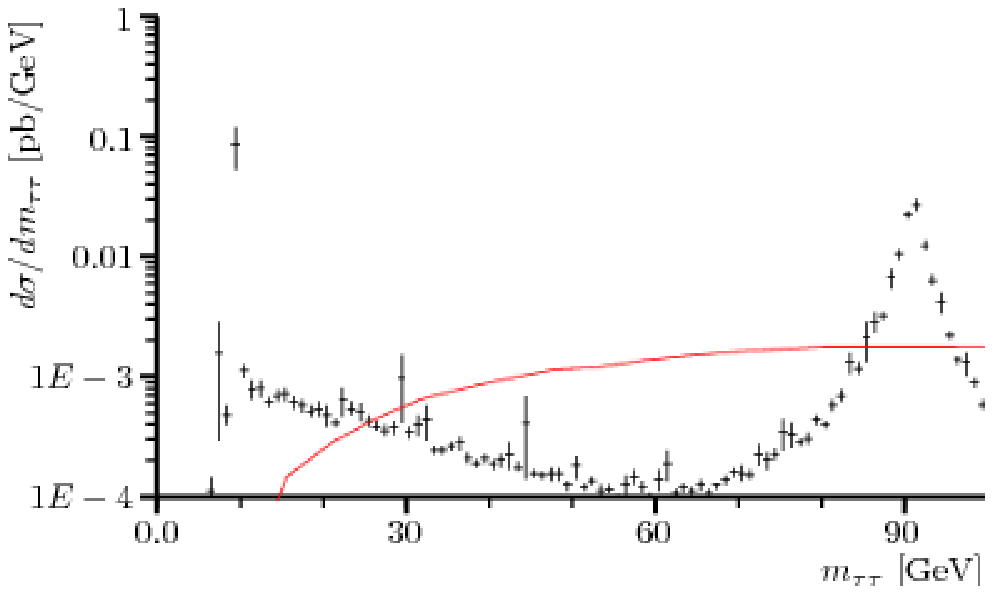}
&\includegraphics[scale=0.75]{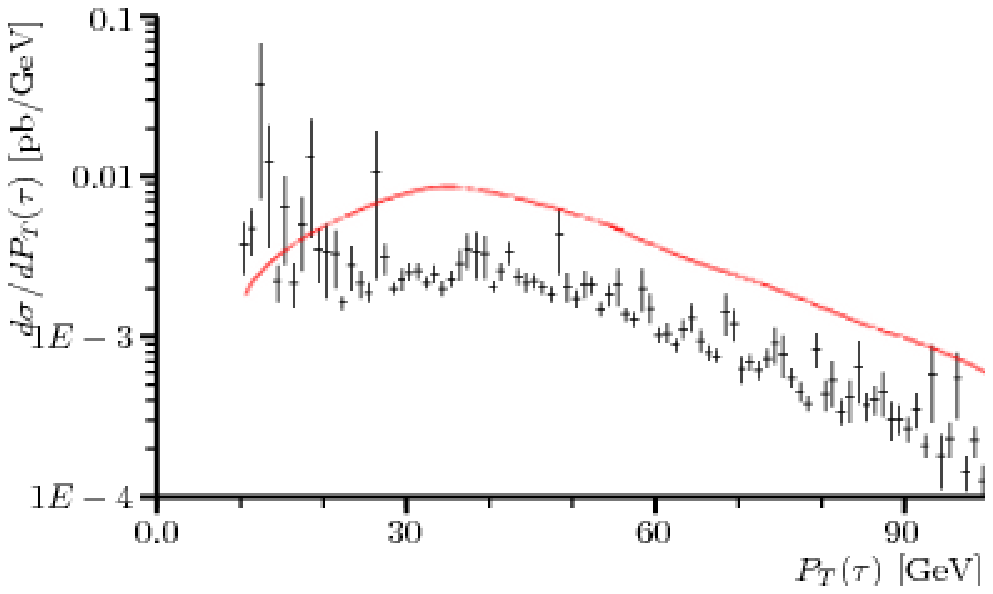}
  \end{tabular} 
\label{fig:TT1}  
\caption{The differential cross section in the $\tau^+\tau^-$ channel for $m_{a_1}$=9.8 GeV as a function of
 $m_{\tau\tau}$ (left) and $P_{T}(\tau)$ (right), after the cuts in (\ref{cuts:TT}). The histogram points represent the signal and 
irreducible background together whereas the red line is $t\bar t$ background.} 
 
\end{figure}

\begin{figure}
 \centering\begin{tabular}{cc}

 \includegraphics[scale=0.75]{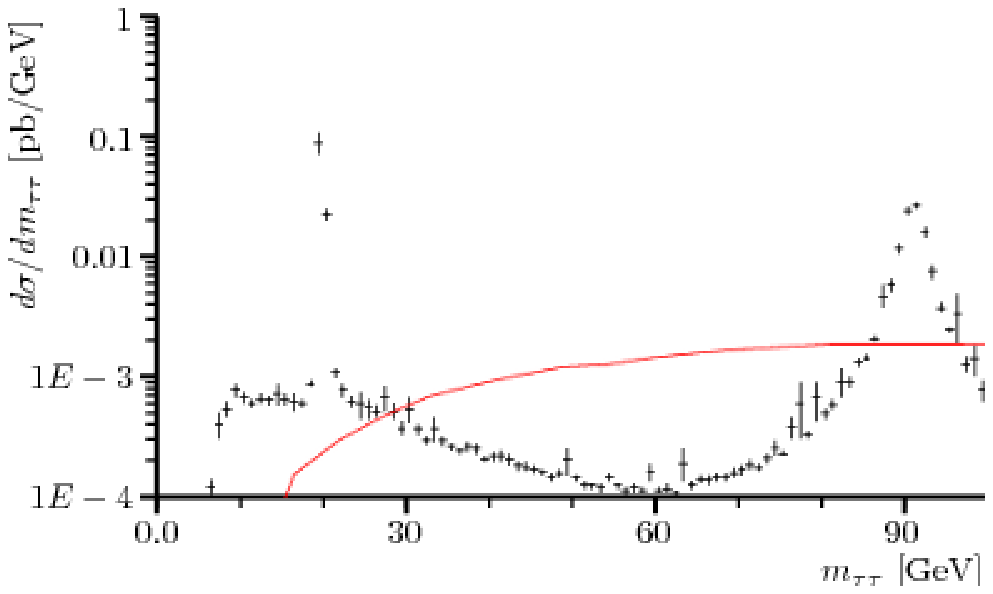}
&\includegraphics[scale=0.75]{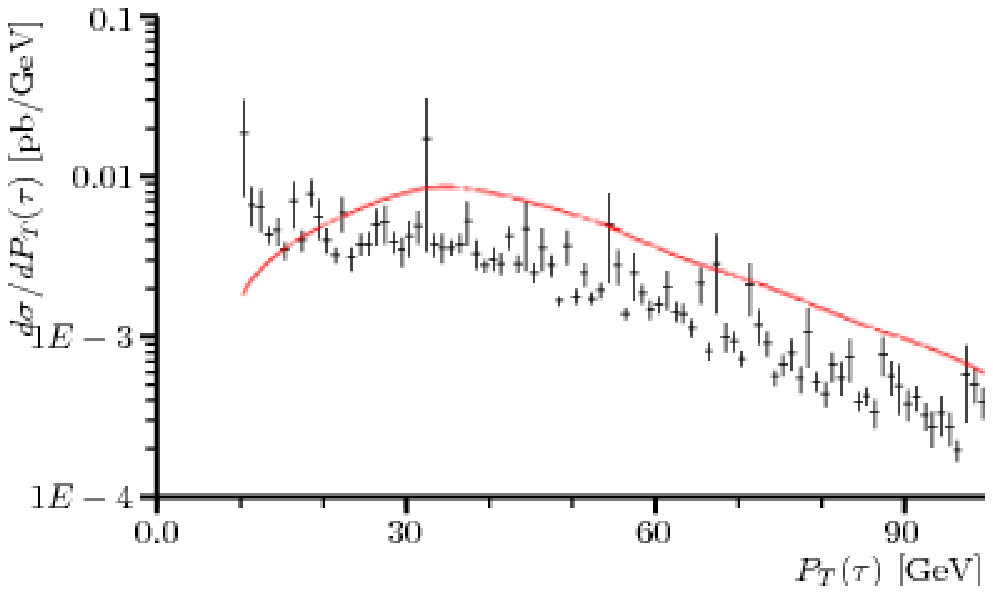}
 \end{tabular}  
\label{fig:TT2}
\caption{The differential cross section in the $\tau^+\tau^-$ channel for $m_{a_1}$=20 GeV 
as a function of  $m_{\tau\tau}$ (left) and $P_{T}(\tau)$ (right), after the cuts in (\ref{cuts:TT}). The histogram points represent the signal 
and irreducible background together whereas the red line is $t\bar t$ background.}
  
\end{figure}

\begin{figure}
 \centering\begin{tabular}{cc}
 \includegraphics[scale=0.75]{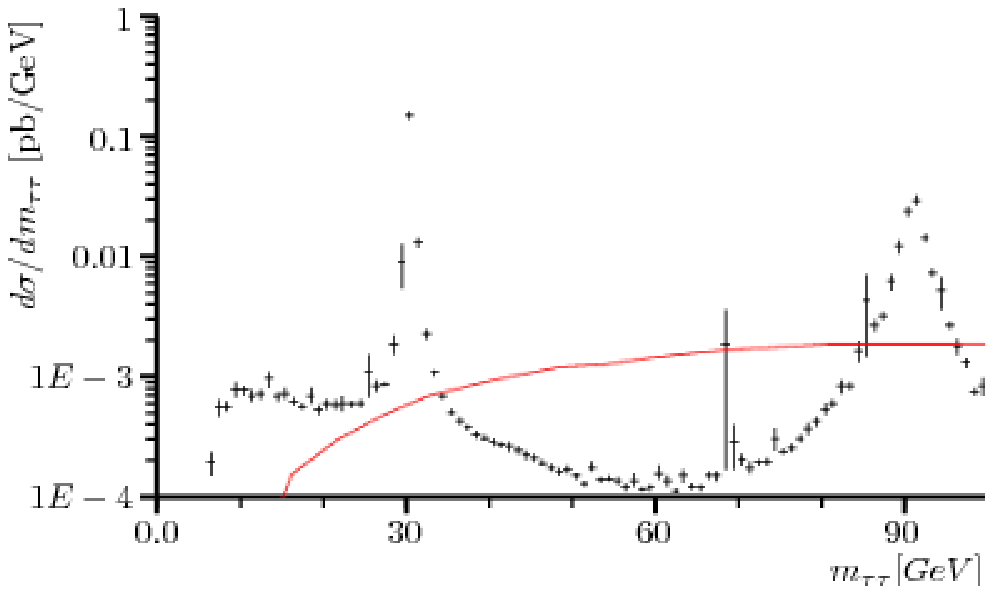}
&\includegraphics[scale=0.75]{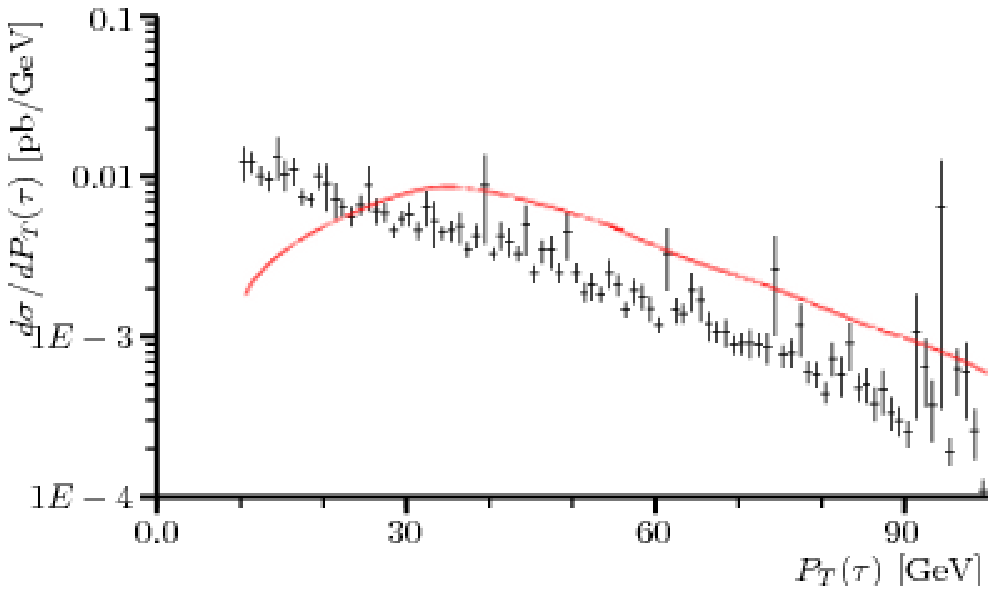}
   \end{tabular} 
\label{fig:TT3}
  \caption{The differential cross section in the $\tau^+\tau^-$ channel for $m_{a_1}$=31 GeV as a function of
  $m_{\tau\tau}$ (left) and $P_{T}(\tau)$ (right), after the cuts in (\ref{cuts:TT}). The histogram points represent the signal and irreducible
 background together whereas the red line is $t\bar t$ background.}
 
\end{figure}

\begin{figure}
 \centering\begin{tabular}{cc}
            
 \includegraphics[scale=0.75]{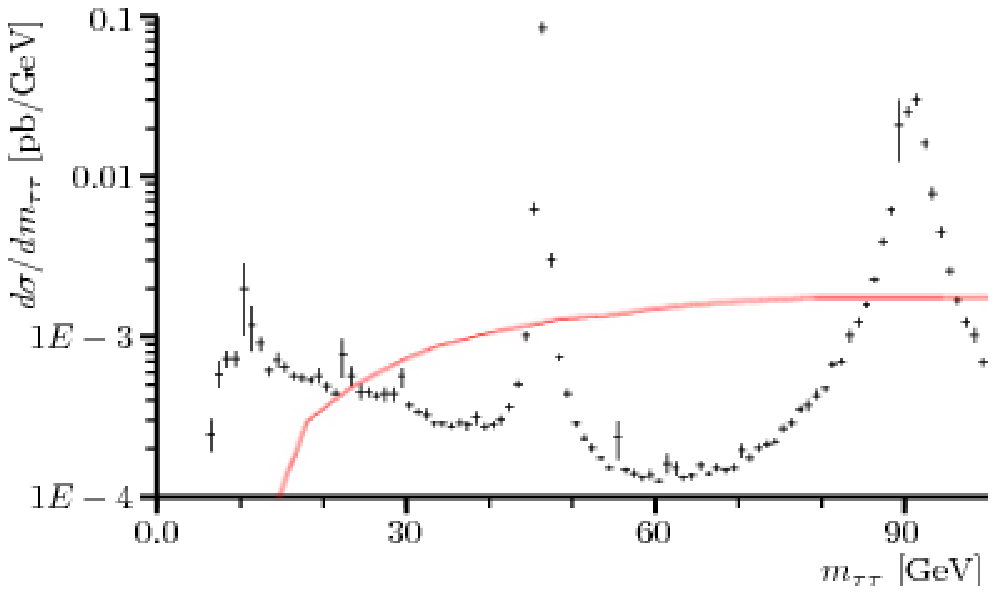}
&\includegraphics[scale=0.75]{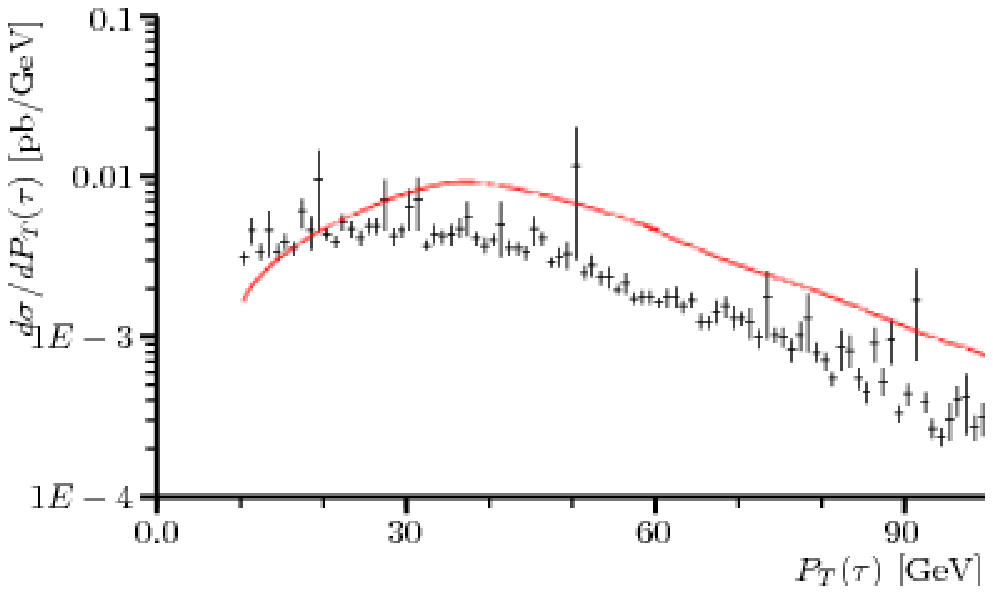}
  \end{tabular}
\label{fig:TT4}  
 \caption{The differential cross section in the $\tau^+\tau^-$ channel for $m_{a_1}$=46 GeV as a function of
 $m_{\tau\tau}$ (left) and $P_{T}(\tau)$ (right), after the cuts in (\ref{cuts:TT}). The histogram points represent the signal and irreducible
 background together whereas the red line is $t\bar t$ background.}

\end{figure}

\begin{figure}
 \centering\begin{tabular}{cc}
 \includegraphics[scale=0.75]{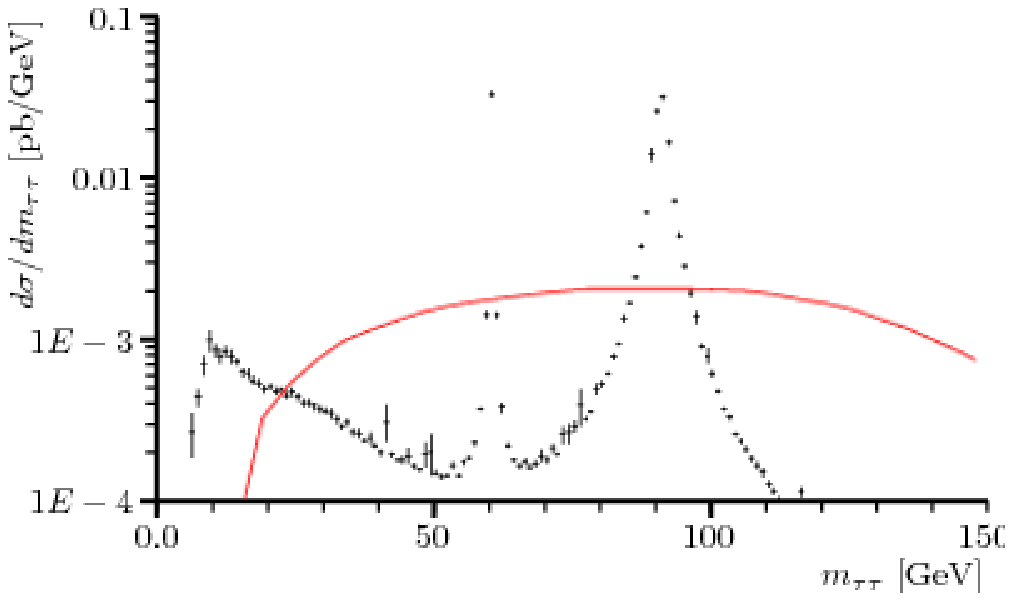}
&\includegraphics[scale=0.75]{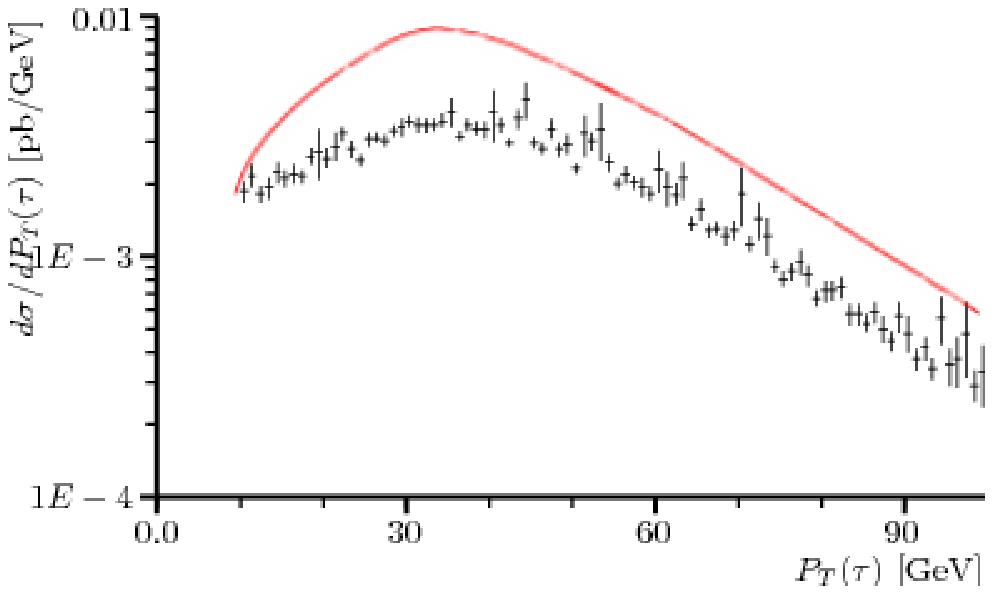}
   \end{tabular}
\label{fig:TT5} 
 \caption{The differential cross section in the $\tau^+\tau^-$ channel for $m_{a_1}$=60.5 GeV as a function of 
$m_{\tau\tau}$ (left) and $P_{T}(\tau)$ (right), after the cuts in (\ref{cuts:TT}). The histogram points represent the signal and irreducible
 background together whereas the red line is $t\bar t$ background.}
  
\end{figure}

\begin{figure}
 \centering\begin{tabular}{cc}
 \includegraphics[scale=0.75]{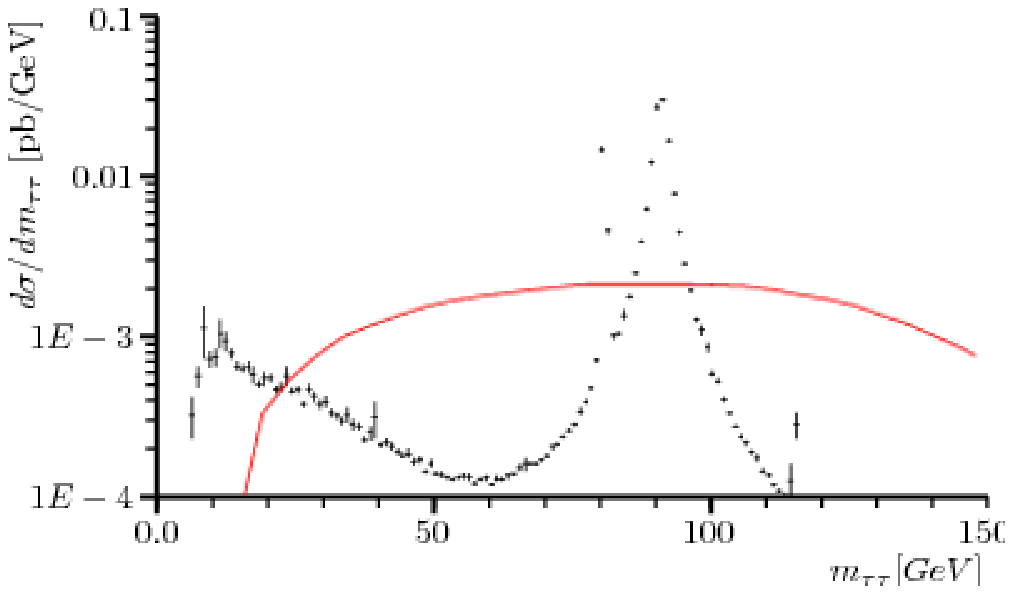}
&\includegraphics[scale=0.75]{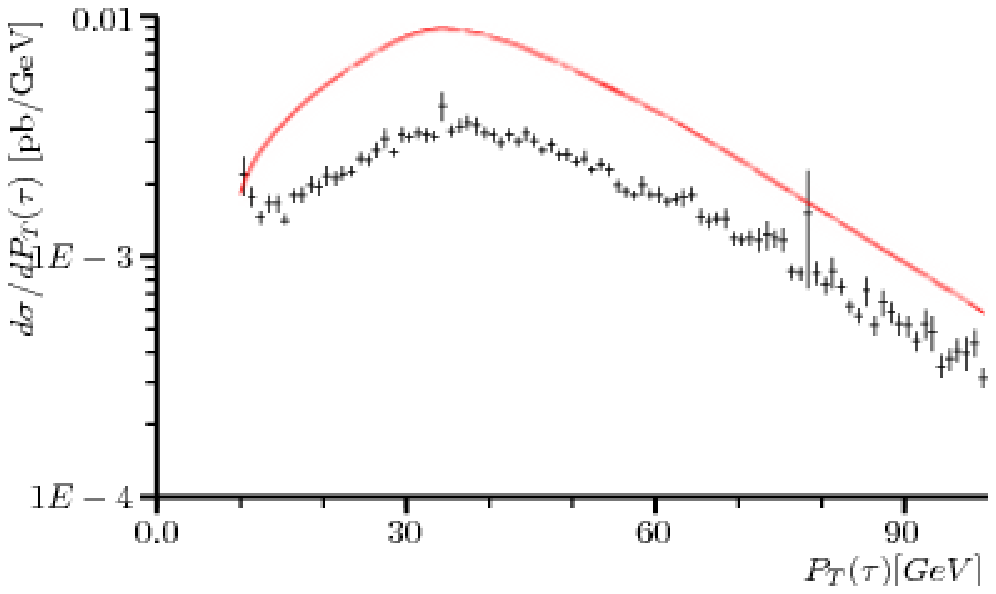}
   \end{tabular}  
\label{fig:TT6}
\caption{The differential cross section in the $\tau^+\tau^-$ channel for $m_{a_1}$=81 GeV as a function of
 $m_{\tau\tau}$ (left) and $P_{T}(\tau)$ (right), after the cuts in (\ref{cuts:TT}). The histogram points represent the signal and irreducible
 background together whereas the red line is $t\bar t$ background.}
 
\end{figure}


\begin{thebibliography}{99}

\bibitem{review} For reviews, see: e.g., U.~Ellwanger, C.~Hugonie and A.~M.~Teixeira, Phys.\ Rept.\ {\bf 496} (2010) 1 (and references therein);
 M.~Maniatis, Int.\ J.\ Mod.\ Phys.\ A {\bf 25} (2010) 3505 (and references therein). 

\bibitem{MNZ}
D.J. Miller, R. Nevzorov and P.M. Zerwas, Nucl. Phys. B \textbf{681} (2004) 3.

\bibitem{upper}  M. Masip, R. Mu${\tilde{\rm n}}$oz-Tapia and A. Pomarol,
Phys. Rev. D {\bf 57} (1998) 5340.

\bibitem{Cyril} U.~Ellwanger and C.~Hugonie,
Eur.\ Phys.\ J.\ C {\bf 25} (2002) 297.

\bibitem{LEPescape} R.~Dermisek and J.~F.~Gunion,
Phys.\ Rev.\ Lett. {\bf 95} (2005) 041801.

\bibitem{NMSSM-Points} U.~Ellwanger, J.~F.~Gunion and C.~Hugonie, JHEP {\bf 0507} (2005) 041;
 U. Ellwanger, J.F. Gunion, C. Hugonie and S. Moretti, hep-ph/0305109 and hep-ph/0401228.

\bibitem{NMSSM-Benchmarks} A.~Djouadi {\it et al.},
  JHEP {\bf 0807} (2008) 002.

\bibitem{NoLoseMSSM}J. Dai, J.F. Gunion, R. Vega, Phys. Lett. B \textbf{315}  (1993) 355 and Phys. Lett. B \textbf{345} (1995)  29;
J.R. Espinosa, J.F. Gunion, Phys. Rev. Lett. \textbf{82}  (1999) 1084.

\bibitem{NoLoseNMSSM1} U.~Ellwanger, J.F.~Gunion and C.~Hugonie, hep-ph/0111179;
D.J. Miller and S. Moretti, hep-ph/0403137; C. Hugonie and S. Moretti, hep-ph/0110241; 
A.~Belyaev, S.~Hesselbach, S.~Lehti, S.~Moretti, A.~Nikitenko and C.~H.~Shepherd-Themistocleous, 
arXiv:0805.3505 [hep-ph]; J.~R.~Forshaw, J.~F.~Gunion, L.~Hodgkinson, A.~Papaefstathiou and 
A.~D.~Pilkington, JHEP {\bf 0804} (2008) 090; A.~Belyaev, J.~Pivarski, A.~Safonov, S.~Senkin and 
A.~Tatarinov, Phys.\ Rev.\ D {\bf 81} (2010) 075021. 

\bibitem{Shobig1} S.~Moretti and S.~Munir,  Eur.\ Phys.\ J.\ C {\bf 47} (2006) 791. 

\bibitem{CPNSH} E.~Accomando {\it et al.}, 
arXiv:hep-ph/0608079.

\bibitem{Shobig2} S.~Moretti, S.~Munir and P.~Poulose,  Phys.\ Lett.\ B {\bf 644} (2007) 241.

\bibitem{Erice} S. Munir, talk given at the `International School of Subnuclear Physics, 43rd Course', 
Erice, Italy, August 29 -- Sept. 7, 2005, to be published in the proceedings, preprint SHEP-05-37,
October 2005.


\bibitem{dirk} D. Zerwas and S. Baffioni, private communication;
S. Baffioni, talk presented at ``GdR Supersym\'etrie 2004, 
5-7 July 2004, Clermont-Ferrand, France.

\bibitem{NMHDECAY}U. Ellwanger, J.F. Gunion and C. Hugonie, JHEP {\bf 0502} (2005) 066;
 U. Ellwanger and C. Hugonie, Comput.\ Phys.\ Commun.\ {\bf 175} (2006) 290.

\bibitem{NMSSMTools} See http://www.th.u-psud.fr/NMHDECAY/nmssmtools.html. 

\bibitem{LEP} S. Schael et al., Eur. Phys. J. C {\bf 47} (2006) 547.


\bibitem{excess} R.~Dermisek and J.~F.~Gunion,  Phys.\ Rev.\ D {\bf 76} (2007) 095006. 



\bibitem{CalcHEP} A.~Pukhov,  arXiv:hep-ph/0412191.

\bibitem{cteq} See http://hep.pa.msu.edu/cteq/public/cteq6.html.

\bibitem{Almarashi} M. M. Almarashi, private programs.
\bibitem{Lebedev} S. Andreas, O. Lebedev, S. R. Sanchez and A. Ringwald, JHEP {\bf 1008} (2010) 003.

\bibitem{bbHiggs-lowmass} See, e.g.: F. Sarri, preprint ATL-PHYS-PROC-2008-076 (and references therein);
S. Horvat,  preprint ATL-PHYS-PROC-2009-063 (and references therein).

\bibitem{ATLAS-TDR} ATLAS Collaboration, arXiv:0901.0512 [hep-ex].

\bibitem{CMS-TDR} CMS Collaboration, J. Phys. G {\bf 34}  (2007) 995.




\end{thebibliography}
\end{document}